\documentclass[a4paper,11pt]{article}

\usepackage{a4wide}
\usepackage{graphicx}
\usepackage{color}
\usepackage{version}
\usepackage[colorlinks,citecolor=blue]{hyperref}
\usepackage[top=1.7cm,bottom=2.5cm,left=1.9cm,right=1.8cm]{geometry}

\usepackage{microtype}

\usepackage{DynamicalIndependence}
\usepackage{mathtools}
\usepackage{amsthm}
\usepackage{tikz-cd}

\usepackage{caption}
\usepackage{subcaption}

\usepackage{lmodern}
\usepackage[T1]{fontenc}

\graphicspath{{figs/}}

\newcommand{\todo}[1][] {\textcolor{red}{$\bigstar$} \textbf{\textcolor{red}{TODO}} \textit{#1}}
\newcommand{\todol}[1][] {\ \\\noindent\todo{#1} \\}

\newcommand{\etc}[0]     {etc.}
\newcommand{\eg}[0]      {e.g.}
\newcommand{\ie}[0]      {i.e.}
\newcommand{\etal}[0]    {\textit{et al.}}
\newcommand{\cf}[0]      {\textit{cf.}}

\newcommand{\gcop}			{\boldsymbol{\mathsf{F}}}
\newcommand{\teop}			{\boldsymbol{\mathsf{T}}}
\newcommand{\miop}			{\boldsymbol{\mathsf{I}}}
\newcommand{\enop}			{\boldsymbol{\mathsf{H}}}

\newcommand{\sgcop}			{\boldsymbol{\mathsf{f}}}
\newcommand\cgraphn         {\mathsf{G}}
\newcommand\mnul            {\square}

\newcommand\trop			{\mathrm{\scalebox{0.65}{\textsf{T}}}}
\newcommand\rrop			{\mathrm{\scalebox{0.65}{\textsf{R}}}}

\DeclareMathOperator{\eig}  {eig}

\newcommand\envi            {E}

\newcommand\sstate          {\mathcal{S}}
\newcommand\estate          {\mathcal{E}}
\newcommand\xstate          {\mathcal{X}}
\newcommand\ystate          {\mathcal{Y}}
\newcommand\zstate          {\mathcal{Z}}
\newcommand\ustate          {\mathcal{U}}
\newcommand\vstate          {\mathcal{V}}

\newcommand\Gmann 			{\mathcal G}
\newcommand\Stief 			{\mathcal V}
\newcommand\Ogroup 			{\mathcal O}

\newcommand\eps				{\varepsilon}
\renewcommand\phi			{\varphi}

\newcommand{\bx}			{\boldsymbol x}
\newcommand{\by}			{\boldsymbol y}

\newcommand{\bu}			{\boldsymbol u}

\newcommand{\bM}			{\boldsymbol M}

\newcommand{\beps}			{\boldsymbol\eps}

\newcommand{\tbeps}			{\tilde\beps}
\newcommand{\tSigma}		{\tilde\Sigma}

\newcommand{\tA}			{\tilde{A}}
\newcommand{\tC}			{\tilde{C}}
\newcommand{\tK}			{\tilde{K}}

\newcommand{\tQ}			{\tilde{Q}}
\newcommand{\tR}			{\tilde{R}}

\newcommand{\tH}			{\tilde{H}}
\newcommand{\tS}			{\tilde{S}}

\newcommand{\tX}			{\tilde{X}}

\renewcommand\prob[1]		    {\probop\!\bracs{#1}}
\renewcommand\expect[1]		{\expectop\!\bracs{#1}}
\renewcommand\trace[1]		{\traceop\!\bracs{#1}}
\renewcommand\var[1]		    {\varop\!\bracs{#1}}

\newcommand\ibar            {:}
\newcommand\vbar            {\,|\,}

\renewcommand\eqref[1]     {(\ref{#1})}

\newcommand\secref[1]      {Section~\ref{#1}}
\newcommand\apxref[1]      {Appendix~\ref{#1}}
\newcommand\figref[1]      {Fig.~\ref{#1}}
\newcommand\figreff[2]     {Figs.~\ref{#1},\,\ref{#2}}

\usepackage{natbib}

\newcommand{\footremember}[2]{%
    \footnote{#2}
    \newcounter{#1}
    \setcounter{#1}{\value{footnote}}%
}
\newcommand{\footrecall}[1]{%
    \footnotemark[\value{#1}]%
}

\numberwithin{equation}{section}

\title{Dynamical independence: discovering emergent macroscopic processes in complex dynamical systems}

\author{%
L. Barnett\footremember{sackler}{Sackler Centre for Consciousness Science, Department of Informatics, University of Sussex, Falmer, Brighton, UK}%
\footremember{corrauth}{Corresponding author: \texttt{\href{mailto:l.c.barnett@sussex.ac.uk}{l.c.barnett@sussex.ac.uk}}\vspace{0.5em}}%
\ and A. K. Seth\footrecall{sackler}
}

\begin{document}

\date\today

\maketitle

\begin{abstract}
We introduce a notion of emergence for coarse-grained macroscopic variables associated with highly-multivariate microscopic dynamical processes, in the context of a coupled dynamical environment. \emph{Dynamical independence} instantiates the intuition of an emergent macroscopic process as one possessing the characteristics of a dynamical system ``in its own right'', with its own dynamical laws distinct from those of the underlying microscopic dynamics. We quantify (departure from) dynamical independence by a transformation-invariant Shannon information-based measure of \emph{dynamical dependence}. We emphasise the data-driven \emph{discovery} of dynamically-independent macroscopic variables, and introduce the idea of a multiscale ``emergence portrait'' for complex systems. We show how dynamical dependence may be computed explicitly for linear systems via state-space modelling, in both time and frequency domains, facilitating discovery of emergent phenomena at all spatiotemporal scales. We discuss application of the state-space operationalisation to inference of the emergence portrait for neural systems from neurophysiological time-series data. We also examine dynamical independence for discrete- and continuous-time deterministic dynamics, with potential application to Hamiltonian mechanics and classical complex systems such as flocking and cellular automata.
\end{abstract}

\section{Introduction} \label{sec:intro}

When we observe a large murmuration of starlings twisting, stretching and wheeling in the dusk, it is hard to escape the impression that we are witnessing an individuated dynamical entity quite distinct from the thousands of individual birds which we know to constitute the flock. The singular dynamics of the murmuration as a whole, it seems, in some sense ``emerges'' from the collective behaviour of its constituents \citep{CavagnaEtal:2013}. Analogously, the gliders and particles observed in some cellular automata appear to emerge as distinct and distinctive dynamical entities from the collective interactions between cells \citep{Bays:2009}. In both cases, these emergent phenomena reveal dynamical structure at coarser ``macroscopic'' scales than the ``microscopic'' scale of interactivity between individual components of the system -- structure which is not readily apparent from the microscopic perspective. Frequently, dynamical interactions at the microscopic level are reasonably simple and/or well-understood; yet an appropriate macroscopic perspective reveals dynamics that do not flow transparently from the micro-level interactions, and, furthermore, appear to be governed by laws quite distinct from the microscopic dynamics. Emergence, it seems, proffers a window into inherent parsimonious structure, across spatiotemporal scales, for a class of complex systems.

In both of the above examples emergent structure ``jumps out at us'' visually. But this need not be the case, and in general may not be the case. For example, directly observing the population activity of large numbers of cortical neurons [\eg, via calcium or optogenetic imaging \citep{WeisenburgerEtal:2019}] may not reveal any visually obvious macroscopic patterning (besides phenomena such as widespread synchrony), even though this activity underlies complex organism-level cognition and behaviour. Even in flocking starlings, while distal visual observation manifestly reveals emergent macroscopic structure, could there be additional emergent structure that would only be apparent from a very different---possibly non-visual---perspective?

In this paper, we address two key questions regarding emergent properties in complex dynamical systems:
\begin{subequations}
\begin{align}
    \parbox{0.88\textwidth}{\textbullet\ How may we \emph{characterise} those perspectives which reveal emergent dynamical structure?} \label{txt:Q1} \\[1em]
    \parbox{0.88\textwidth}{\textbullet\ Knowing the microscopic dynamics, how may we \emph{find} these revealing perspectives?} \label{txt:Q2}
\end{align} \label{txt:QQ}%
\end{subequations}
By providing principled data-driven methods for identifying emergent structure across spatiotemporal scales, we hope to enable new insights into many complex systems from brains to ecologies to societies.

\subsection{Emergence and dynamical independence} \label{sec:emergence}

Emergence is broadly understood as a gross (macroscopic) property of a system of interacting elements, which is not a property of the individual (microscopic) elements themselves. A distinction is commonly drawn between between ``strong'' and ``weak'' emergence \citep{Bedau:1997}. A strongly-emergent macroscopic property (i) is in principle not deducible from its microscopic components, and (ii) has irreducible causal power over these components. This flavour of emergence appears to reject mechanistic explanations altogether, and raises awkward metaphysical issues about causality, such as how to resolve competition between competing micro- and macro-level ``downward'' causes \citep{Kim:2006}. By contrast, \citet[p.~375]{Bedau:1997} characterises emergent phenomena as ``somehow constituted by, and generated from, underlying processes'', while at the same time ``somehow autonomous from underlying processes''. He goes on to define a process as weakly emergent iff it ``can be derived from [micro-level dynamics and] external conditions but only by simulation''. Weakly emergent properties are therefore ontologically reducible to their microscopic causes, though they remain epistemically opaque from these causes.

We propose a new notion and measure of emergence inspired by Bedau's formulation of weak emergence. Our notion, \emph{dynamical independence}, shares with weak emergence the aim to capture the sense in which a flock of starlings seems to have a ``life of its own'' distinct from the microscopic process (interactions among individual birds), even though there is no mystery that the flock is in fact constituted by the birds. Following Bedau's formulation, a dynamically-independent macroscopic process is ``ontologically reducible'' to its microscopic causes, and downward (physical) causality is precluded. However, dynamically-independent macroscopic processes may display varying degrees of ``epistemic opacity'' from their microscopic causes, loosening the constraint that (weak) emergence relations can only be understood through exhaustive simulation.

Dynamical independence is defined for macroscopic dynamical phenomena associated with a microscopic dynamical system---macroscopic variables---which \emph{supervene}\footnote{A property A is said to be supervenient on a property B if A ``depends on'' B, in the sense that a difference in the state of A implies a difference in the state of B.} \citep{Davidson:1980} on the microscopic. Here, supervenience of macro on micro is operationalised in a looser predictive sense: that macroscopic variables convey no information about their own evolution in time beyond that conveyed by the microscopic dynamics (and possibly a coupled environment). The paradigmatic example of such macroscopic variables, and one which we mostly confine ourselves to in this study, is represented by the \emph{coarse-graining} of the microscopic system by aggregation of microscopic components at some characteristic scale. Dynamical independence is framed in predictive terms: a macroscopic variable is defined to be dynamically-independent if, even while supervenient on the microscopic process, knowledge of the microscopic process adds nothing to prediction of the macroscopic process beyond the extent to which the macroscopic process already self-predicts. (This should not be taken to imply, however, that a dynamically-independent process need self-predict \emph{well}, if indeed at all; see discussion in \secref{sec:discussion}.)

To bolster intuition, consider a large group of particles, such as a galaxy of stars. The system state is described by the ensemble of position and momentum vectors of the individual stars in some inertial coordinate system, and the dynamics by Newtonian gravitation. We may construct a low-dimensional coarse-grained macroscopic variable by taking the average position and total momentum (and, if we like, also the angular momentum and total energy) of the stars in the galaxy. Elementary physics tells us that this macroscopic variable in fact self-predicts perfectly without any knowledge of the detailed microscopic state; it has a ``life of its own'', perfectly understandable without recourse to the microscopic level. Yet an arbitrarily-concocted coarse-graining---\ie, an arbitrary mapping of the microscopic state space to a lower-dimensional space---will almost certainly \emph{not} have this property\footnote{The macroscopic variables in this example are, of course, \emph{conserved} (time-invariant) quantities, and as such trivially self-predicting; see \secref{sec:flow} for further discussion.}: indeed, the vast majority of coarse-grainings do not define dynamically-independent processes.

Dynamical independence is defined over the range of scales from microscopic, through mesoscopic to macroscopic. It is expressed, and quantified, solely in terms of Shannon (conditional) mutual information \citep{CoverThomas:1991}, and as such is fully \emph{transformation-invariant}; that is, for a physical process it yields the same quantitative answers no matter how the process is measured. Under some circumstances, it may be defined in the frequency domain, thus enabling analysis of emergence across temporal scales. It applies in principle  to a broad range of dynamical systems, continuous and discrete in time and/or state, deterministic and stochastic. Examples of interest include Hamiltonian dynamics, linear stochastic systems, neural systems, cellular automata, flocking, econometric processes, and evolutionary processes.

As previously indicated, our specific aims are \eqref{txt:Q1} to \emph{quantify} the degree of dynamical independence of a macroscopic variable, and \eqref{txt:Q2} given the micro-level dynamics, to \emph{discover} dynamically-independent macroscopic variables. In the current article we address these aims primarily for stochastic processes in discrete time, and analyse in detail the important and non-trivial case of stationary linear systems.

The article is organised as follows: in \secref{sec:approach} we set out our approach. We present the formal underpinnings of dynamical systems, macroscopic variables and coarse-graining, the information-theoretic operationalisation of dynamical independence, its quantification and its properties. We declare an \emph{ansatz} on a practical approach to our primary objectives \eqref{txt:QQ}. In \secref{sec:cglss} we specialise to linear stochastic systems in discrete time, and analyse in depth how our \emph{ansatz} may be achieved for linear state-space systems; in particular, we detail how dynamically-independent macroscopic variables may be discovered in state-space systems via numerical optimisation, and present a worked example illustrating the procedure. In \secref{sec:detcon} we discuss approaches to dynamical independence for deterministic and continuous-time systems.  In \secref{sec:discussion} we summarise our findings, discuss related approaches in the literature and examine some potential applications in neuroscience.

\section{Approach} \label{sec:approach}

\subsection{Dynamical systems} \label{sec:dysys}

Our notion of dynamical independence applies to dynamical systems. We describe a dynamical system by a sequence of variables $S_t$ taking values in some state space $\sstate$ at times indexed by $t$ (when considered as a whole, we sometimes drop the time index and write just $S$ to denote the sequence $\{S_t\}$). In full generality, the state space might be discrete or real-valued, and possibly endowed with further structure (\eg, topological, metric, linear, \etc). The sequential time index may be discrete or continuous. The dynamical law, governing how $S_t$ evolves over time, specifies the system state at time $t$ given the history of states at  prior times $t' < t$; this specification may be deterministic or probabilistic. In this study, we largely confine our attention to discrete-time stochastic processes, where the $S_t$, $t \in \integers$, are jointly-distributed random variables. The distribution of $S_t$ is thus contingent on previously-instantiated historical states; that is, on the set $s_t^- = \{s_{t'} : t' < t\}$ given that $S_{t'} = s_{t'}$ for $t' < t$ (throughout this article we use a superscript dash to denote sets of prior states).  In \secref{sec:detcon}, we discuss extension of dynamical independence to deterministic and/or continuous-time systems.

The general scenario we address is a ``dynamical universe'' that may be partitioned into a ``microscopic'' dynamical system of interest $X_t$ coupled to a dynamical environment $E_t$, where $X_t$ and  $E_t$ are jointly-stochastic variables taking values in state spaces $\xstate$ and $\estate$ respectively. Typically, the microscopic state will be the high-dimensional ensemble state of a large number of atomic elements, \eg, birds in a flock, molecules in a gas, cells in a cellular automaton, neurons in a neural system, \etc\ The microscopic and environmental\footnote{Compared with other approaches to emergence in the literature (\secref{sec:relapp}), the environment takes a back seat in our current treatment; here our primary concern is the microscopic/macroscopic relationship (\secref{sec:macrovar}).} processes jointly constitute a dynamical system $(X_t,E_t)$; that is, the dynamical laws governing the evolution in time of microscopic and environmental variables will depend on their \emph{joint} history. Here we assume that the system/environment boundary is given [see \citet{KrakauerEtal:2020} for an approach to the challenge of distinguishing systems from their environments].

\subsection{Macroscopic variables and coarse-graining} \label{sec:macrovar}

Given a microscopic dynamical system $X_t$ and environment $E_t$, we associate an emergent phenomenon explicitly with some ``macroscopic variable'' associated with the microscopic system. Intuitively, we may think of a macroscopic variable as a gross perspective on the system, a ``way of looking at it'' (\cf~\secref{sec:intro}), or a particular mode of description of the system \citep{ShaliziMoore:2003,AllefeldAtmanspacherWackermann:2009}. We operationalise this idea in terms of a process $Y_t$ that in some sense aggregates microscopic states in $\xstate$ into common states in a lower-dimension or cardinality state space $\ystate$, with a consequent loss of information. (We don't rule out that aggregation may occur over time as well as state.) The dimension or cardinality of $\ystate$ defines the \emph{scale}, or ``granularity'', of the macroscopic variable.

The supervenience of macroscopic variables on the microscopic dynamics (\secref{sec:emergence}) is operationalised in predictive, information-theoretic terms: we assume that a macroscopic variable conveys no information about its own future beyond that conveyed by the joint microscopic and environmental histories. Explicitly, we demand the condition
\begin{equation}
	\miop(Y_t \ibar Y_t^- \vbar X_t^-,E_t^-) \equiv 0 \label{eq:macrovar}
\end{equation}
where $\miop(\cdot\ibar\cdot\vbar\cdot)$ denotes Shannon conditional mutual information\footnote{Throughout, if the state space is continuous then mutual information is defined in terms of \emph{differential} entropy \citep{CoverThomas:1991}.} \citep{CoverThomas:1991}. A canonical example of a macroscopic variable in the above sense is one of the form $Y_t = f(X_t)$, where $f \ibar \xstate \to \ystate$ is a deterministic, surjective mapping from the microscopic onto the lower dimensional/cardinality macroscopic state space (here aggregation is over states, but not over time\footnote{This might be extended to variables of the form $Y_t = f(X_t,X_t^-)$, which aggregate over time as well as state; the condition \eqref{eq:macrovar} still holds.}). The relation \eqref{eq:macrovar} is then trivially satisfied. We refer to this as \emph{coarse-graining}, to be taken in the broad sense of dimensionality reduction. Coarse-graining partitions the state space, ``lumping together'' microstates in the preimage $f^{-1}(y) \subseteq \xstate$ of macrostates $y \in \ystate$, with a concomitant loss of information: many microstates correspond to the same macrostate. For concision, we sometimes write $Y = f(X)$ to denote the coarse-graining $Y_t = f(X_t)$.

If the state space $\xstate$ is endowed with some structure (\eg, topological, metric, smooth, linear, \etc) then we generally restrict attention to structure-preserving mappings (morphisms). In particular, we restrict coarse-grainings $f$ to epimorphisms (surjective structure-preserving mappings)\footnote{We might frame our analysis more formally in the language of Category Theory \citep{MacLane:1978}.}. There is a natural equivalence relation amongst coarse grainings: given $f \ibar \xstate \to \ystate$, $f' \ibar \xstate \to \ystate'$ we write
\begin{equation}
	f' \sim f \iff \exists \text{ an isomorphism } \psi \ibar \ystate \to \ystate' \text{ such that } f' = \psi \circ f \label{eq:fequiv}
\end{equation}
$f$ and $f'$ then lump together the same subsets of microstates. When we talk of a coarse-graining, we implicitly intend an equivalence class $\{f\}$ of mappings $\xstate \to \ystate$ under the equivalence relation \eqref{eq:fequiv}. In the remainder of this article we restrict attention to coarse-grained macroscopic variables.

\subsection{Dynamical independence} \label{sec:DI}

As we have noted, not \emph{every} coarse-graining $f : \xstate \to \ystate$ will yield a macroscopic variable $Y_t = f\big(X_t\big)$ which we would be inclined to describe as emergent (\cf~the galaxy example in \secref{sec:emergence}). Quite the contrary; for a complex microscopic system comprising many interacting components, we may expect that an arbitrary coarse-grained macroscopic variable will fail to behave as a dynamical entity in its own right, with its own distinctive law of evolution in time. When applied to the coarse-grained variable, the response to the question: \emph{What will it do next?} will be: \emph{Well, without knowing the full \emph{microscopic} history, we really can't be sure}; unsurprising, perhaps, as coarse-graining, by construction, loses information.

By contrast, for an \emph{emergent} macroscopic variable, despite the loss of information incurred by coarse-graining, the macroscopic dynamics are parsimonious in the following sense: knowledge of the microscopic history adds nothing to the capacity of the macroscopic variable to self-predict. Dynamical independence formalises this parsimony as follow  (we temporarily disregard the environment):
\begin{equation}
\parbox{0.79\textwidth}{
	Given jointly-stochastic processes $(X,Y)$, $Y$ is dynamically-independent of $X$ iff, conditional on its own history, $Y$ is independent of the history of $X$.
} \label{eq:dians1}
\end{equation}
In information-theoretic terms, \eqref{eq:dians1} holds (at time $t$) precisely when $\miop(Y_t \ibar X_t^- \vbar Y_t^-)$ vanishes identically. We recognise this quantity as the \emph{transfer entropy} \citep[TE;][]{Schreiber:2000,Palus:2001,KaiserSchreiber:2002,BossomaierEtal:2016} $\teop_t(X \to Y)$ from $X$ to $Y$ at time $t$. Thus we state formally:
\begin{equation}
\parbox{0.79\textwidth}{
	$Y$ is dynamically-independent of  $X$ at time $t \iff \teop_t(X \to Y) \equiv 0$.
} \label{eq:dians2}
\end{equation}
Eq.~\eqref{eq:dians2} establishes an information-theoretic \emph{condition}\footnote{In the language of \citet{BertschingerEtal:2006}, $Y$ is ``informationally closed'' with respect to $X$; see \secref{sec:relapp}.} for dynamical independence of $Y$ with respect to $X$; we further propose the transfer entropy $\teop_t(X \to Y)$, the \emph{dynamical dependence} of $Y$ on $X$, as a quantitative, non-negative measure of the extent to which $Y$ \emph{departs} from dynamical independence with respect to $X$ at time $t$. Crucially, ``dynamical independence''  refers to the condition, while ``dynamical dependence'' refers to the measure\footnote{This is entirely analogous to statistical independence, a condition on the relationship between jointly distributed random variables, as opposed to mutual information, a quantitative measure of statistical dependence.}.

Dynamical (in)dependence is naturally interpreted in predictive terms: the \emph{un}predictability of the process $Y$ at time $t$ given its own history is naturally quantified by the entropy rate $\enop(Y_t \vbar Y_t^-)$. We may contrast this with the unpredictability $\enop(Y_t \vbar X_t^-,Y_t^-)$ of $Y$ given not only its own history, but also the history of $X$. Thus the dynamical dependence $\teop_t(X \to Y) = \enop(Y_t \vbar Y_t^-) - \enop(Y_t \vbar Y_t^-,X_t^-)$ quantifies the extent to which $X$ predicts $Y$ over-and-above the extent to which $Y$ already self-predicts.

As presented above, dynamical independence generalises straightforwardly to take account of a third, jointly-stochastic conditioning variable, via conditional transfer entropy \citep{BossomaierEtal:2016}. In the case where $X$ represents a microscopic system, $E$ a  jointly-stochastic environmental process and $Y$ an associated macroscopic variable, we define our dynamical dependence measure as
\begin{subequations}
\begin{align}
	\teop_t(X \to Y \vbar \envi)
	&= \miop(Y_t \ibar X_t^- \vbar Y_t^-,\envi_t^-) \label{eq:tedda} \\
	&= \enop(Y_t \vbar Y_t^-,E_t^-) - \enop(Y_t \vbar X_t^-,Y_t^-,E_t^-) \label{eq:teddb} \\
	&= \enop(Y_t \vbar Y_t^-,E_t^-) - \enop(Y_t \vbar X_t^-,E_t^-) \qquad\qquad\text{from \eqref{eq:macrovar}} \label{eq:teddc}
\end{align} \label{eq:tedd}%
\end{subequations}
and define the condition for dynamical independence as:
\begin{equation}
\parbox{0.79\textwidth}{
	The macroscopic variable $Y$ is \emph{dynamically-independent} of the microscopic system $X$ in the context of the environment $E$ at time $t$ $\iff$ $\teop_t(X \to Y \vbar \envi) \equiv 0$.
} \label{eq:didef}
\end{equation}
The dynamical dependence $\teop_t(X \to Y \vbar \envi)$ will in general be time-varying, except when all processes are strongly-stationary; for the remainder of this article we restrict ourselves to the stationary case and drop the time index subscript. In the case that the processes $X,Y,E$ are deterministic, mutual information is not well-defined, and dynamical independence must be framed differently. We discuss deterministic systems in \secref{sec:detcon}.

As a conditional Shannon mutual information \citep{CoverThomas:1991}, the transfer entropy $\teop(X \to Y \vbar \envi)$ is \emph{nonparametric} in the sense that it is invariant with respect to reparametrisation of the target, source and conditional variables by isomorphisms of the respective state spaces \citep{KaiserSchreiber:2002}. Thus if $\phi$ is an isomorphism of $\xstate$, $\psi$ an isomorphism of $\ystate$ and $\chi$ an isomorphism of $\estate$, then
\begin{equation}
	\teop(\phi(X) \to \psi(Y) \vbar \chi(\envi)) \equiv \teop(X \to Y \vbar \envi) \label{eq:ddinvar}
\end{equation}
In particular, dynamical (in)dependence respects the equivalence relation \eqref{eq:fequiv} for coarse-grainings. This means that (at least for coarse-grained macroscopic variables) transfer entropy from macro to micro vanishes trivially; \ie, $\teop(Y \to X \vbar \envi) \equiv 0$, which we may interpret  as the non-existence of ``downward causation''.

To guarantee \emph{transitivity} of dynamical independence (see below), we introduce a mild technical restriction on admissible coarse-grainings to those $f : \xstate \to \ystate$ with the following property:
\begin{equation}
	\exists\; \text{an epimorphism}\ u : \xstate \to \ustate \ \text{such that} \ \phi = f \times u : \xstate \to \ystate \times \ustate \ \text{is an isomorphism} \label{eq:fsplit}
\end{equation}
Intuitively, there is a ``complementary'' mapping $u$ which, along with $f$ itself, defines a nonsingular transformation of the system. For example, if $f$ is a projection of the real Euclidean space $\reals^n$ onto the first $m < n$ coordinates, $u$ could be taken as the complementary projection of $\reals^n$ onto the remaining $n-m$ coordinates (\cf~\secref{sec:cglss}). Trivially, \eqref{eq:fsplit} respects the equivalence relation \eqref{eq:fequiv}. The restriction holds universally for some important classes of structured dynamical systems, \eg, the linear systems analysed in \secref{sec:cglss}, and also in general for discrete-state systems; otherwise, it might be relaxed to obtain at least ``locally'' in state space\footnote{For some systems (\eg, where the only structure is set-theoretic and the state space uncountable) \eqref{eq:fsplit} may require the Axiom of Choice \citep{SEP-AOC:2015}.}. We assume \eqref{eq:fsplit} for all coarse-grainings from now on.

Given property \eqref{eq:fsplit}, we may apply the transformation $\phi = f \times u$ and exploit the dynamical dependence invariance \eqref{eq:ddinvar} to obtain an equivalent system $X \sim (Y,U)$, $U = u(X)$ in which the coarse-graining $Y$ becomes a projection of $\xstate = \ystate \times \ustate$ onto $\ystate$, and dynamical dependence is given by
\begin{equation}
	\teop(X \to Y \vbar \envi) = \teop(Y,U \to Y \vbar \envi) = \teop(U \to Y \vbar \envi) \label{eq:displit}
\end{equation}
Assuming \eqref{eq:fsplit} for all coarse-grainings, using \eqref{eq:displit} we may show that dynamical independence is transitive:
\begin{equation}
\parbox{0.79\textwidth}{
	In the context of the environmental process $E$, if $Y = f(X)$ is dynamically-independent of $X$ and $Z = g(Y)$ is dynamically-independent of $Y$, then $Z = (g \circ f)(X)$ is dynamically-independent of $X$
} \label{eq:ditrans}%
\end{equation}
We provide a formal proof in \apxref{sec:ditrans}. We have thus a partial ordering on the set of coarse-grained dynamically-independent macroscopic variables, under which they may potentially be hierarchically nested at increasingly coarse scales.

Systems featuring emergent properties are typically large ensembles of dynamically interacting elements; that is, system states $x \in \xstate$ are of the form $(x_1,\ldots,x_n) \in \xstate_1 \times \ldots \times \xstate_n$ with $\xstate_k$ the state space of the $k$th element, where $n$ is large. Dynamical independence for such systems may be related to the \emph{causal graph} of the system \citep{Barnett:mvgc:2014,Seth:gcneuro:2015}, which encapsulates information transfer between system elements. As we shall see (below and \secref{sec:disim}), this facilitates the construction of systems with prescribed dynamical-independence structure. Given a coarse-graining $y_k = f_k(x_1,\ldots,x_n)$, $k = 1,\ldots,m$ at scale $m$, using \eqref{eq:fsplit} it is not hard to show that we may always transform the system so that $y_k = x_k$ for $k = 1,\ldots,m$; that is, under some ``change of coordinates'', the coarse-graining becomes a projection onto the subspace defined by the first $m$ dimensions of the microscopic state space. The dynamical dependence is then given by
\begin{equation}
	\teop(X \to Y \vbar \envi) = \teop(X_{m+1},\ldots,X_n \to X_1,\ldots,X_m \vbar \envi) \label{eq:tesplit}
\end{equation}
and we may show\footnote{A proof may be constructed along the same lines as the proof in \apxref{sec:ditrans}.} that, under such a transformation
\begin{equation}
	\teop(X \to Y \vbar \envi) = 0 \iff \cgraphn_{ij}(X \vbar \envi) = 0 \quad\text{for}\quad i = 1,\ldots,m\,, \quad j = m+1,\ldots,n \label{eq:dicg}
\end{equation}
where
\begin{equation}
	\cgraphn_{ij}(X \vbar \envi) = \teop\big(X_j \to X_i \vbar X_{[ij]},\envi\big)\\, \qquad i,j = 1,\ldots,n\,,\quad i \ne j \label{eq:tecgraph}
\end{equation}
is the causal graph of the system $X$ conditioned on the environment (here the subscript ``$[ij]$'' denotes omission of the $i$ and $j$ components of $X$). According to \eqref{eq:dicg} we may characterise dynamically-independent macroscopic variables for ensemble systems as those coarse-grainings which are transformable into projections onto a sub-graph of the causal graph with no incoming information transfer from the rest of the system. However, given two or more dynamically-independent macroscopic variables (at the same or different scales), in general we cannot expect to find a transformation under which all of those variables \emph{simultaneously} become projections onto causal sub-graphs. Nonetheless, \eqref{eq:dicg} is useful for constructing dynamical systems with prespecified dynamically-independent macroscopic variables (\cf~\secref{sec:disim}).

For many complex dynamical systems, there will be no fully dynamically-independent macroscopic variables at some particular (or perhaps at any) scale; \ie, no macroscopic variables for which \eqref{eq:didef} holds \emph{exactly}\footnote{When we wish to stress that the condition \eqref{eq:didef} is satisfied exactly, we shall sometimes refer to ``perfect'' dynamical independence (\cf~\secref{sec:cdi}).}. There may, however, be macroscopic variables for which the dynamical dependence \eqref{eq:tedd}  is small -- in an empirical scenario, for instance, ``small'' might be defined as statistically insignificant. We take the view that even ``near-dynamical independence'' yields useful structural insights into emergence, and adopt the \emph{ansatz}:
\begin{subequations}
\begin{align}
    \parbox{0.79\textwidth}{\textbullet\ The \emph{maximally} dynamically-independent macroscopic variables at a given scale (\ie, those which minimise dynamical dependence) characterise  emergence at that scale.} \label{txt:ansatz1} \\[1em]
    \parbox{0.79\textwidth}{\textbullet\ The collection of maximally dynamically-independent macroscopic variables at \emph{all} scales, along with their degree of dynamical dependence, affords a multiscale portrait of the emergence structure of the system.} \label{txt:ansatz2}
\end{align} \label{txt:ansatz}%
\end{subequations}

\section{Linear systems} \label{sec:cglss}

In this section we consider linear discrete-time continuous-state systems, and later specialise to linear state-space (SS) systems. For simplicity, we consider the case of an empty environment, although the approach is readily extendable to the more general case.

Our starting point is that the microscopic system is, or may be modelled as, a wide-sense stationary, purely-nondeterministic\footnote{That is, the deterministic (perfectly predictable) component of the Wold moving-average decomposition  of the process \citep{Doob:1953} is identically zero.}, stable,  minimum-phase\footnote{Minimum-phase requires that the system have a stable inverse \citep{HandD:2012}; see also \citet{Geweke:1982}.} (``miniphase''), zero-mean, vector\footnote{All vectors are considered to be \emph{column} vectors, unless otherwise stated.} stochastic process $X_t = [X_{1t} \ X_{2t} \ \ldots \ X_{nt}]^\trop$, $t \in \integers$ defined on the vector space $\reals^n$. These conditions guarantee that the process has unique stable and causal vector moving-average (VMA) and vector autoregressive (VAR) representations:
\begin{subequations}
\begin{align}
	X_t = \beps_t + \sum_{k=1}^\infty B_k \beps_{t-k} & \quad\text{or}\quad X_t = H(z) \cdot \beps_t\,, && \ H(z) = I + \sum_{k=1}^\infty B_k z^k \label{eq:MA} \\
	&\text{and} \notag \\
	X_t = \sum_{k=1}^\infty A_k X_{t-k} +\beps_t & \quad\text{or}\quad H(z)^{-1} \cdot X_t = \beps_t\,, && \ H(z)^{-1} = I - \sum_{k=1}^\infty A_k z^k \label{eq:AR}
\end{align} \label{eq:MAAR}%
\end{subequations}
respectively, where $\beps_t$ is a white noise (serially uncorrelated, iid) innovations process with covariance matrix $\Sigma = \expectop[\beps_t\beps_t^\trop]$, and $H(z)$ the transfer function with $z$ the back-shift operator (in the frequency domain, $z = e^{-i\omega}$, with $\omega$ = angular frequency in radians). $B_k$ and $A_k$ are respectively the VMA and VAR coefficient matrices. The stability and miniphase conditions imply that both the $B_k$ and $A_k$ are square-summable, and that all zeros and poles of the transfer function lie strictly outside the unit disc in the complex plane\footnote{The \emph{spectral radius} of the process is given by  $\rho = \max\{|1/z| : H(z)^{-1} = 0\}$; stability requires that $\rho < 1$.}. The cross-power spectral density (CPSD) matrix for the process is given by \citep{Wilson:1972}
\begin{equation}
	S(z) = H(z) \Sigma H^*(z) \label{eq:cpsd}
\end{equation}
where ``*'' denotes conjugate transpose. Note that at this stage we do not assume that the innovations $\beps_t$ are (multivariate) Gaussian. Note too, that even though we describe the system \eqref{eq:MAAR} as ``linear'', \emph{this does not necessarily exclude processes with nonlinear generative mechanisms} -- we just require that the conditions listed above are met. Wold's Decomposition Theorem \citep{Doob:1953} guarantees a VMA form \eqref{eq:MA} provided that the process is wide-sense stationary and purely-nondeterministic; if in addition the process is miniphase, the VAR form \eqref{eq:AR} also exists, and all our conditions are satisfied. Thus our analysis here also covers a large class of stationary ``nonlinear'' systems, with the caveats that (i) for a given nonlinear generative model, the VMA/VAR representations will generally be infinite-order, and as such may not represent parsimonious models for the system, and (ii) restriction of coarse-graining to the linear domain (see below) may limit analysis to macroscopic variables which lack a natural relationship with the nonlinear structure of the dynamics. Nonetheless, linear models are commonly deployed in a variety of real-world scenarios, especially for econometric and neuroscientific time-series analysis. Reasons for their popularity include parsimony (linear models will frequently have fewer parameters than alternative nonlinear models\footnote{This is particularly pertinent in econometric time-series analysis, where typically low signal-to-noise ratios heavily penalise complex models.}), simplicity of estimation, and mathematical tractability.

Since we are in the linear domain, we restrict ourselves to linear coarse-grained macroscopic variables (\cf~\secref{sec:macrovar}). A surjective linear mapping $L : \reals^n \to \reals^m$, $0 < m < n$, corresponds to a full-rank $m \times n$ matrix, and the coarse-graining equivalence relation \eqref{eq:fequiv} identifies $L,L'$ iff there is a non-singular linear transformation $\Psi$ of $\reals^m$ such that $L' = \Psi L$. (Note that since $L$ is full-rank, $Y_t = LX_t$ is purely-nondeterministic and satisfies all the requirements listed at the beginning of this section.) Considering the rows of $L$ as basis vectors for an $m$-dimensional linear subspace of $\reals^n$, a linear transformation  simply specifies a change of basis for the subspace. Thus we may identify the set of linear coarse-grainings with the \emph{Grassmannian manifold} $\Gmann_m(n)$ of $m$-dimensional linear subspaces of $\reals^n$. The Grassmannian \citep{Helgason:1978} is a compact smooth manifold of dimension $m(n-m)$. It is also a non-singular \emph{algebraic variety} (the set of solutions of a system of polynomial equations over the real numbers), a \emph{homogeneous space} (it ``looks the same at any point''), and an \emph{isotropic space} (it ``looks the same in all directions''); specifically, $\Gmann_m(n) = \Ogroup(n)/\big(\Ogroup(m) \times \Ogroup(n-m)\big)$, where $\Ogroup(n)$ is the Lie group of real orthogonal matrices. Under the Euclidean inner-product (vector dot-product), every $m$-dimensional subspace of $\reals^n$ has a unique orthogonal complement of dimension $n-m$\footnote{An orthonormal basis for the orthogonal complement may be found using a Singular Value Decomposition (SVD).}, which establishes a (non-canonical) isometry of $\Gmann_m(n)$ with $\Gmann_{n-m}(n)$; for instance, in $\reals^3$, every line through the origin has a unique orthogonal plane through the origin, and vice-versa. There is a natural definition of \emph{principal angles} between linear subspaces of Euclidean spaces,
via which the Grassmannian $\Gmann_m(n)$ may be endowed with various invariant metric structures \citep{Wong:1967}.

By transformation-invariance of dynamical dependence, we may assume without loss of generality that the row-vectors of $L$ form an orthonormal basis; \ie,
\begin{equation}
	LL^\trop = I \label{eq:Lnorm}
\end{equation}
The manifold of linear mappings satisfying \eqref{eq:Lnorm} is known as the \emph{Stiefel manifold} $\Stief_m(n) \equiv \Ogroup(n)/\Ogroup(n-m)$, which, like the Grassmannian is a compact, homogeneous and isotropic algebraic variety, with dimension $nm-\tfrac12m(m+1)$. In contrast to the set of all full-rank mappings $\reals^n \to \reals^m$, the Stiefel manifold is bounded, which is advantageous for computational minimisation of dynamical dependence (\secref{sec:macglssn}).

The condition \eqref{eq:fsplit} is automatically satisfied for linear coarse-grainings. In particular, given $L$ satisfying \eqref{eq:Lnorm}, we may always find a surjective linear mapping $M : \reals^n \to \reals^{n-m}$ where the row-vectors of the $(n-m) \times n$ matrix $M$ form an orthonormal basis for the orthogonal complement of the subspace spanned by the row-vectors of $L$. The transformation
\begin{equation}
	\Phi = \begin{bmatrix} L \\ M \end{bmatrix} \label{eq:LMxform}
\end{equation}
of $\reals^n$ is then nonsingular and orthonormal; \ie, $\Phi\Phi^\trop = I$.

Given a linear mapping $L$, our task is to calculate the dynamical dependence $\teop(X \to Y)$ for the coarse-grained macroscopic variable $Y_t = LX_t$. In the context of linear systems, it is convenient to switch from transfer entropy to Granger causality \citep[GC;][]{Wiener:1956,Granger:1963,Granger:1969,Geweke:1982}. In case the innovations $\beps_t$ in \eqref{eq:MAAR} are multivariate-normal, the equivalence of TE and GC is exact \citep{Barnett:tegc:2009}; else we may either consider the GC approach as an approximation to ``actual'' dynamical dependence\footnote{The measures are in fact equivalent under a somewhat broader class of distributions \citep{HlavackovaSchindler:2011}, and more generally asymptotically equivalent under a Gaussian likelihood approximation \citep{Barnett:teml:2012}.}, or, if we wish, consider dynamical dependence framed in terms of GC rather than TE as a linear prediction-based measure in its own right; we note that key properties of dynamical (in)dependence including transformation invariance \eqref{eq:ddinvar}, the existence of complementary mappings \eqref{eq:fsplit} [\cf~\eqref{eq:LMxform}], transitivity \eqref{eq:ditrans} and relationship to the (Granger-)causal graph \eqref{eq:dicg} carry over straightforwardly to the GC case. GC has distinct advantages over TE in terms of analytic tractability, sample estimation and statistical inference, in both parametric \citep[\cf~\secref{sec:lsss} below]{Barnett:mvgc:2014,Barnett:ssgc:2015} and nonparametric \citep{Dhamala:2008a} scenarios. In \apxref{sec:gc} we provide a concise recap of (unconditional) Granger causality following the classical formulation of \citet{Geweke:1982}.

\subsection{Linear state-space systems} \label{sec:lsss}

We now specialise to the class of linear state-space systems \eqref{eq:MAAR} (under the restrictions listed at the beginning of \secref{sec:cglss}), where $X_t$ may be represented by a model of the form
\begin{subequations}
\begin{align}
	W_{t+1} &= AW_t + U_t && \text{state-transition equation,} \label{eq:ssx} \\
	X_t     &= CW_t + V_t && \text{observation equation,} \label{eq:ssy}
\end{align} \label{eq:ss}%
\end{subequations}
where the (unobserved) state process $W_t = [W_{1t} \ W_{2t} \ \ldots \ W_{rt}]^\trop$, $t \in \integers$, is defined on $\reals^r$, $U_t,V_t$ are zero-mean multivariate white noises, $C$ is the observation matrix and $A$ the state transition matrix. Note the specialised use of the term ``state space'' in the linear systems vocabulary: the state variable $W_t$ is to be considered a notional unobserved process, or simply as a mathematical construct for expressing the dynamics of the observation process $X_t$, which here stands as the ``microscopic variable''.

The parameters of the model \eqref{eq:ss} are $(A,C,Q,R,S)$, where
\begin{equation}
	\begin{bmatrix} Q & S \\ S^\trop & R  \end{bmatrix} \equiv
	\expect{ \hspace{-1pt}
	\begin{bmatrix} U_t \\ V_t \end{bmatrix} \hspace{-6pt}
 	\begin{array}{c}
		\begin{bmatrix} U_t^\trop & \!\! V_t^\trop \end{bmatrix} \\ \phantom.
	\end{array} \hspace{-4pt}
	}
\end{equation}
is the joint noise covariance matrix (the purely-nondeterministic assumption implies that $R$ is positive-definite). Stationarity requires that the transition equation \eqref{eq:ssx} satisfy the stability condition $\max\{|\lambda| : \lambda \in \eig(A)\} < 1$. A process $X_t$ satisfying a stable, miniphase SS model \eqref{eq:ss} also satisfies a stable, miniphase vector autoregressive moving-average (VARMA) model; conversely, any stable, miniphase VARMA process satisfies a stable, miniphase SS model of the form \eqref{eq:ss} \citep{HandD:2012}.

To facilitate calculation of dynamical dependence (\secref{sec:acglss} below), it is useful to transform the SS model \eqref{eq:ss} to ``innovations form'' (\cf~\apxref{sec:ssgc})
\begin{subequations}
\begin{align}
	Z_{t+1} &= AZ_t + K\beps_t \label{eq:issz} \\
	X_t     &= CZ_t + \phantom{K}\beps_t \label{eq:issy}
\end{align} \label{eq:iss}%
\end{subequations}
with new state variable $Z_t = \cexpect{W_t}{X^-_t}$, white-noise \emph{innovations} process $\beps_t = X_t - \cexpect{X_t}{X^-_t}$ with covariance matrix $\Sigma = \expect{\beps\beps^\trop}$, and \emph{Kalman gain} matrix $K$. The moving-average and autoregressive operators for the innovations-form state-space (ISS) model \eqref{eq:iss} are given by
\begin{subequations}
\begin{align}
	H(z) &= I + C(1-Az)^{-1}Kz \label{eq:ssMA} \\
	&\text{and} \notag \\
	H(z)^{-1} &= I - C(1-Bz)^{-1}Kz \label{eq:ssAR}
\end{align} \label{eq:ssMAAR}%
\end{subequations}
respectively, where $B = A-KC$. The miniphase condition is thus $\max\{|\lambda| : \lambda \in \eig(B)\} < 1$.

A general-form SS \eqref{eq:ss} may be converted to an ISS \eqref{eq:iss} by solving the associated discrete algebraic Riccati equation \citep[DARE;][]{LancasterRodman:1995,HandD:2012}.
\begin{equation}
	P =  APA^\trop + Q - \big(APC^\trop + S\big) \big(CPC^\trop + R\big)^{-1} \big(CPA^\trop + S^\trop\big) \label{eq:dare}
\end{equation}
which under our assumptions has a  unique stabilising solution for $P$; then
\begin{subequations}
\begin{align}
	\Sigma &= CPC^\trop + R \label{eq:p2iSig} \\
	K &= \big(APC^\trop +S\big) \Sigma^{-1} \label{eq:p2iK}
\end{align} \label{eq:p2i}%
\end{subequations}

\subsection{Dynamical dependence for state-space systems} \label{sec:acglss}

From \eqref{eq:ss} it is clear that a macroscopic process $Y_t = LX_t$ will be of the same form; that is, the class of state-space systems is closed under full-rank linear mappings. Now consider the (nonsingular) orthonormal transformation \eqref{eq:LMxform} above. Setting $\tX_t = \Phi X_t$, again by transformation invariance we  have $\gcop(X \to Y) = \gcop(\tX \to Y)$, and $\tX_t$ satisfies the ISS model
\begin{subequations}
\begin{align}
	Z_{t+1} &= AZ_t + \tK \tbeps_t \label{eq:tissz} \\
	\tX_t     &= \tC Z_t + \phantom{\tK}\tbeps_t \label{eq:tissy}
\end{align} \label{eq:tiss}%
\end{subequations}
where $\tbeps_t = \Phi\beps_t$ so that $\tSigma = \Phi\Sigma\Phi^\trop$, $\tC = \Phi C$ and $\tK = K\Phi^\trop$ (note that by orthonormality, $\Phi^{-1} = \Phi^\trop$). Now partitioning $\tX_t$ into $\tX_{1t} = LX_t = Y_t$ and $\tX_{2t} = MX_t$, by transformation-invariance we have $\gcop(\tX \to Y) = \gcop(\tX \to \tX_1) = \gcop(\tX_2 \to \tX_1)$, where the last equality holds since, given $\tX_{1t}^-$, the all-variable history $\tX_t^-$ yields no additional predictive information about $\tX_{1t}$ beyond that contained in $\tX_{2t}^-$.

We may now apply the recipe for calculating (unconditional) GC for an innovations-form SS system, as described in \apxref{sec:ssgc}, to
\begin{equation}
	\gcop(X \to Y) = \gcop(\tX_2 \to \tX_1) = \log\frac{\big|\tSigma_{11}^\rrop\big|}{\big|\tSigma_{11} \big|}
\end{equation}
We find $\tC_1 = LC$ and \eqref{eq:rrescovs} becomes
\begin{equation}
	\tQ = K \Sigma K^\trop\,, \qquad
	\tR = L \Sigma L^\trop\,, \qquad
	\tS = K  \Sigma L^\trop \label{eq:rrescovst}
\end{equation}
The DARE \eqref{eq:rdare} for calculating the innovations-form parameters for the reduced model for $\tX_{1t}$ then becomes
\begin{equation}
	P =	APA^\trop + K \Sigma K^\trop - \big(A P C^\trop + K\Sigma\big) L^\trop \bracs{L\big(CPC^\trop + \Sigma\big)L^\trop}^{-1} L\big(A P C^\trop + K\Sigma)^\trop \label{eq:dddare}
\end{equation}
which has a unique stabilising solution for $P$. Explicitly, setting
\begin{equation}
	V = CPC^\trop + \Sigma
\end{equation}
we have $\tSigma_{11}^\rrop = LVL^\trop$, so that finally
\begin{equation}
	\gcop(X \to Y) = \log\frac{|LVL^\trop|}{|L \Sigma L^\trop|} \label{eq:autonom}
\end{equation}
Note that $P$ and $V$ are implicitly functions of the ISS parameters $(A,C,K,\Sigma)$ and the matrix $L$.

Again using transformation invariance, we note that transformation of $\reals^n$ by the inverse of the left Cholesky factor of $\Sigma$ yields $\tSigma = I$ in the transformed system. Thus from now on, without loss of generality we restrict ourselves to the case $\Sigma = I$, as well as the orthonormalisation \eqref{eq:Lnorm}. This further simplifies the DARE \eqref{eq:dddare} to
\begin{equation}
	P =	APA^\trop + K K^\trop - \big(A P C^\trop + K\big) L^\trop \bracs{L\big(CPC^\trop + I\big)L^\trop}^{-1} L\big(A P C^\trop + K)^\trop \label{eq:dddares}
\end{equation}
and the dynamical dependence \eqref{eq:autonom} becomes simply
\begin{equation}
	\gcop(X \to Y) = \log|LVL^\trop|\,, \qquad V = CPC^\trop + I \label{eq:ssdd}
\end{equation}

\subsubsection{Frequency domain} \label{sec:spectral}

Like $\gcop(X \to Y)$, the spectral GC $\sgcop(X \to Y;z)$ (see \apxref{sec:ssgc}) is invariant under nonsingular linear transformation of source or target variable at all frequencies \citep{Barnett:gcfilt:2011}. To calculate the spectral dynamical dependence $\sgcop(X \to Y;z)$, we again apply the orthonormal transformation \eqref{eq:LMxform} and calculate $ \sgcop(X \to Y;z) = \sgcop(\tX_2 \to \tX_1;z)$. Firstly, we may confirm that the transfer function transforms as $\tH(z) = \Phi H(z) \Phi^\trop$, so that by \eqref{eq:cpsd} we have $\tS(z) = \Phi S(z) \Phi^\trop$; in particular, we have $\tH_{12}(z) = LH(z)M^\trop$ and $\tS_{11}(z) = LS(z)L^\trop$. We may then calculate that under the normalisations $LL^\trop = I$ and $\Sigma = I$, we have $\tSigma_{22|1} = I$, so that, noting that $LL^\trop + MM^\trop = I$, we have
\begin{equation}
	\sgcop(X \to Y;z) = \log\frac{\big|LS(z)L^\trop\big|}{\big|LH(z)L^\trop LH^*(z)L^\trop\big|} \label{eq:sdd}
\end{equation}
with $S(z) = H(z)H^*(z)$. As per \eqref{eq:blgc}, we may define the \emph{band-limited dynamical dependence} as
\begin{equation}
	\gcop(X \to Y;\omega_1,\omega_2) = \frac1{\omega_2-\omega_1} \int_{\omega_1}^{\omega_2} \sgcop\big(X \to Y;e^{-i\omega}\big)\,d\omega \label{eq:bldd}
\end{equation}
Noting that $LH(z)L^\trop$ is the transfer function and $LH(z)L^\trop LH^*(z)L^\trop$ the CPSD for the process $Y$, by a standard result \citep[Theorem~4.2]{Rozanov:1967} we have $\frac1{2\pi} \int_0^{2\pi} \log\big|LH\big(e^{-i\omega}\big)L^\trop LH^*\big(e^{-i\omega}\big)L^\trop\big|\,d\omega = \log|L \Sigma L^\trop| = \log|I| = 0$. From \eqref{eq:sdd} and \eqref{eq:sgcint} the time-domain dynamical dependence is thus compactly expressed as
\begin{equation}
	\gcop(X \to Y) = \frac1{2\pi} \int_0^{2\pi} \log\big|LS\big(e^{-i\omega}\big)L^\trop\big|\,d\omega \label{eq:ssgcint}
\end{equation}
which may be more computationally convenient and/or efficient than \eqref{eq:ssdd} with the DARE \eqref{eq:dddares} (\cf~ \secref{sec:macglssn}).

We note that in the presence of an environmental process, we must consider \emph{conditional} spectral GC, which is somewhat more complex than the unconditional version \eqref{eq:sgc} \citep{Geweke:1984,Barnett:ssgc:2015}; we leave this for a future study.

\subsubsection{Finite-order VAR systems} \label{sec:diar}

Finite-order autoregressive systems (\apxref{sec:gcar}) are an important special case of state-space (equivalently VARMA) systems. For a VAR($p$), $p < \infty$, \eqref{eq:ar}, to calculate $\gcop(X \to Y)$ we may convert the VAR($p$) to an equivalent ISS \eqref{eq:ariss} \citep{HandD:2012}, and proceed as in \secref{sec:acglss} above. Alternatively, we may exploit the dimensional reduction detailed in \apxref{sec:gcar}: applying the transformation \eqref{eq:LMxform} with normalisation $LL^\trop = I$, it is easy to calculate that the autoregressive coefficients transform as $\tA_k = \Phi A_k \Phi^\trop$ for $k = 1,\ldots,p$, and as before $\tSigma = \Phi\Sigma\Phi^\trop$. We thus find [\cf~\eqref{eq:A22C12}] $\tA_{22} = \bM A \bM^\trop$ and $\tC_{12} = L C \bM^\trop$, where $\bM = \diag{M,\ldots,M}$ ($p$ blocks of $M$ on the diagonal), and with normalisation $\Sigma = I$, we have [\cf~\eqref{eq:Q22RS21}].
\begin{equation}
	Q_{22} =
	\begin{bmatrix}
		I_{22} & 0 & \ldots & 0 \\
		0      & 0 & \ldots & 0 \\
		\vdots & \vdots & \ddots & \vdots \\
		0      & 0 & \ldots & 0
	\end{bmatrix}\,, \qquad\quad
	R = I_{11}\,, \qquad\quad
	S_{21} = 0
\end{equation}
so that, setting $\Pi = M^\trop P M$ for compactness,
\begin{equation}
	\gcop(X \to Y) = \log|LVL^\trop|\,, \qquad V = C_{12} \Pi C_{12}^\trop + I \label{eq:ardd}
\end{equation}
with $P$ the unique stabilising solution of the reduced $p(n-m) \times p(n-m)$ DARE \eqref{eq:ardare1}
\begin{equation}
		P = \bM\big\{A_{22} \Pi A_{22}\trop + Q_{22} - (A_{22} \Pi C_{12}^\trop) L^\trop \big[L(C_{12} \Pi C_{12}^\trop + I)L^\trop\big]^{-1} L (A_{22} \Pi C_{12}^\trop)^\trop \big\} \bM^\trop \label{eq:ddardare}
\end{equation}
For spectral GC, the formula \eqref{eq:sdd} applies, with transfer function as in \eqref{eq:artf}.

\subsection{Perfect dynamical independence} \label{sec:cdi}

We now examine generic conditions under which perfectly dynamically-independent macroscopic variable can be expected to exist, where by ``generic'' we mean ``except on a measure-zero subset of the model parameter space''. Again applying the transformation \eqref{eq:LMxform}, \eqref{eq:ssgc0} yields
\begin{equation}
	\gcop(X \to Y) \equiv 0 \iff LC A^k KM^\trop \equiv 0 \quad\text{for}\quad k = 0,1,\ldots,r-1 \label{eq:ssdd0}
\end{equation}
where $r$ is the dimension of the state space. Eq.~\eqref{eq:ssdd0} constitutes $rm(n-m)$ multivariate-quadratic equations for the $m(n-m)$ free variables which parametrise the Grassmannian $\Gmann_m(n)$. For $r = 1$, we would thus expect solutions yielding dynamically-independent $Y_t = LX_t$ at all scales $0 < m < n$; however, as the equations are quadratic, some of these solutions may not be real. For $r > 1$, except on a measure-zero subset of the $(A,C,K)$ ISS parameter space, there will be solutions  if $LC \equiv 0$ or $KM^\trop \equiv 0$ (or both). The former comprises $rm$ linear equations, and the latter $r(n-m)$ equations, for the $m(n-m)$ Grassmannian parameters. Therefore, we expect generic solutions to \eqref{eq:ssdd0} if $r < n$ and either $m \le n-r$ or $m \ge r$ (or $r \le m \le n-r$, in which case $2r \le n$ is required). Generically, for $r \ge n$ there will be no perfectly dynamically-independent macroscopic variables. We note that $r < n$ corresponds to ``simple'' models with few spectral peaks; nonetheless, anecdotally it is not uncommon to estimate parsimonious model orders $< n$ for highly multivariate data, especially for limited time-series data.

In the generic VAR($p$) case \eqref{eq:ar}, the condition \eqref{eq:argc0} for vanishing GC \citep{Geweke:1982} yields
\begin{equation}
	\gcop(X \to Y) \equiv 0 \iff LA_kM^\trop \equiv 0 \quad\text{for}\quad k = 1,\ldots,p \label{eq:argc01}
\end{equation}
which constitutes $pm(n-m)$ multivariate-quadratic equations for the $m(n-m)$ Grassmannian parameters. Generically, for $p = 1$ we should again expect to find dynamically-independent macroscopic variables at all scales $0 < m < n$, while for $p > 1$ we don't expect to find any dynamically-independent macroscopic variables, except on a measure-zero subset of the VAR($p$) parameter space $(A_1,\ldots,A_p)$.

Regarding spectral dynamical independence, we note that $\sgcop(X \to Y;z)$ is an \emph{analytic} function of $z = e^{-i\omega}$. Thus by a standard property of analytic functions, if band-limited dynamical independence \eqref{eq:bldd} vanishes for any particular finite interval $[\omega_1,\omega_2]$ then it is zero everywhere, so that by \eqref{eq:sgcint} the time-domain dynamical dependence \eqref{eq:ssdd} must also vanish identically.

\subsection{Statistical inference} \label{sec:statsinf}

Given empirical time-series data, a VAR or state-space model may be estimated via standard (maximum-likelihood) techniques, such as ordinary least squares \citep[OLS;][]{Hamilton:1994} for VAR estimation, or a subspace method \citep{VOandDM:1996} for state-space estimation. The dynamical dependence, as a Granger causality sample statistic, may then in principle be tested for significance at some prespecified level, and dynamical independence of a coarse-graining $Y_t = LX_t$ inferred by failure to reject the null hypothesis of zero dynamical dependence \eqref{eq:ssdd0} [see also \apxref{sec:ssgc}, eq.~\eqref{eq:ssgc0}].

In the case of dynamical dependence calculated from an estimated state-space model [\ie, via \eqref{eq:autonom} or \eqref{eq:ssdd}], the asymptotic null sampling distribution is not known, and surrogate data methods would be required. For VAR modelling, the statistic \eqref{eq:ardd} is a ``single-regression'' Granger causality estimator, for which an asymptotic generalised $\chi^2$ sampling distribution has recently been obtained by \citet{GutknechtBarnett:2019}. Alternatively, a likelihood-ratio, Wald or $F$-test \citep{Lutkepohl:2005} might be performed for the null hypothesis \eqref{eq:argc0} of vanishing dynamical dependence \eqref{eq:argc01}.

\subsection{Maximising dynamical independence} \label{sec:maxdi}

Following our \textit{ansatz} \eqref{txt:ansatz}, given an ISS system \eqref{eq:iss}, whether or not perfectly dynamically-independent macroscopic variables exist at any given scale, we seek to minimise the dynamical dependence $\gcop(X \to Y)$ over the Grassmannian manifold of linear coarse-grainings (\ie, over $L$ for $Y_t = LX_t$). The band-limited dynamical dependence $\gcop(X \to Y;\omega_1,\omega_2)$ \eqref{eq:blgc} may also in principle be minimised at a given scale to yield maximally dynamically-independent coarse-grainings associated with the given frequency range at that scale;  we leave this for future research.

Solving the minimisation of dynamical dependence \eqref{eq:ssdd} over the Grassmannian \emph{analytically} appears, at this stage, intractably complex (see \apxref{sec:macglssa} for a standard approach); we thus proceed to numerical optimisation.

\subsubsection{Numerical optimisation} \label{sec:macglssn}

Given a set of ISS parameters $(A,C,K)$ (we may as before assume $\Sigma = I$), minimising the \emph{cost function} $\gcop(X \to Y)$ of \eqref{eq:ssdd} over the Grassmannian manifold $\Gmann_m(n)$ of linear subspaces presents some challenges. Note that we are not (yet) able to calculate the gradient of the cost function explicitly, provisionally ruling out a large class of gradient-based optimisation techniques\footnote{Implementation of gradient methods on the Grassmannian is, furthermore, decidedly nontrivial \citep{EdelmanEtal:1998}.}.

Simulations (\secref{sec:disim}) indicate that the cost function \eqref{eq:ssdd} appears to be in general multi-modal, so optimisation procedures may tend to find local sub-optima. We do not consider this a drawback; rather, in accordance with our \textit{ansatz} \eqref{txt:ansatz}, we consider them of interest in their own right, as an integral aspect of the emergence portrait.

While $\Gmann_m(n)$ is compact of dimension $m(n-m)$, its parametrisation over the $nm-\tfrac12m(m+1)$-dimensional Stiefel manifold $\Stief_m(n)$ of $m \times n$ orthonormal basis matrices $L$ is many-to-one. There will thus be $\tfrac12m(m-1)$-dimensional equi-cost surfaces in the Stiefel manifold. These zero-gradient sub-manifolds may confound standard optimisation algorithms; population-based (non-gradient) methods such as cross-entropy optimisation \citep{BotevEtal:2013}, for example, fail to converge when parametrised by $\reals^{mn}$ under the constraint \eqref{eq:Lnorm}, apparently because the population diffuses along the equi-cost surfaces. Preliminary investigations suggest that simplex methods \citep{NelderMead:1965}, which are generally better at locating global optima, also fare poorly, although the reasons are less clear.

An alternative approach is to use local coordinate charts for the Grassmannian. Any full-rank $m \times n$ matrix $L$ can be represented as
\begin{equation}
	L = \Psi[I_{m\times m} \; M]\Pi \label{eq:LQM}
\end{equation}
where $\Pi$ is an $n \times n$ permutation matrix (\ie, a row or column permutation of $I_{n\times n}$), $\Psi$ an $m \times m$ non-singular transformation matrix and $M$ is $m \times (n-m)$ full-rank. For given $\Pi$ the Grassmannian is then locally and diffeomorphically mapped by the $m \times (n-m)$ full-rank matrices $M$\footnote{These charts comprise the ``standard atlas'' used to define the canonical differentiable manifold structure on the Grassmannian.}. But note that for given $\Pi,M$, while there is no redundancy in the (injective) mapping $M : \reals^{m(n-m)} \to \Gmann_m(n)$, the space of such $M$ is unbounded and doesn't cover the entire Grassmannian, which again makes numerical optimisation awkward.

A partial resolution is provided by a surprising mathematical result due to \citet{Knuth:1985} [see also \citep{UsevichMarkovsky:2014}], which states roughly that given any fixed $\delta > 1$, for any full-rank $m \times n$ matrix $L_0$ there is a neighbourhood of $L_0$, a permutation matrix $\Pi$, and a transformation $\Psi$ such that for any $L$ in the neighbourhood of $L_0$, all elements of $M$ satisfying \eqref{eq:LQM} are bounded to lie in $[-\delta, \delta]$. That is, in the local neighbourhood of any subspace in the Grassmannian, we can always find a suitable permutation matrix $\Pi$ such that \eqref{eq:LQM} effectively parametrises the neighbourhood by a \emph{bounded} submanifold of $\reals^{m(n-m)}$. During the course of an optimisation process, then, if the current local search (over $M$) drifts outside its $\delta$-bounds, we can always find a \emph{new} bounded local parametrisation of the search neighbourhood ``on the fly''. Finding a suitable new $\Pi$ is, however, not straightforward, and calculating the requisite $\Psi$ for \eqref{eq:LQM} is quite expensive computationally \citep{MehrmannPoloni:2012}. Nor is this scheme particularly convenient for population-based optimisation algorithms, which will generally require keeping track of different permutation matrices for different sub-populations, and---worse---for some algorithms (such as CE optimisation), it seems that the procedure can only work if the entire current population resides in a single $(\Psi,\Pi)$-chart.

Regarding computational efficiency, we may apply a useful pre-optimisation trick. From \eqref{eq:ssdd0}, $\gcop(X \to Y)$ vanishes precisely where the ``proxy cost function''
\begin{equation}
	\gcop^*(X \to Y) = \sum_{k=0}^{r-1} \big\| LC A^k KM^\trop \big\|^2
\end{equation}
vanishes, where as before $M$ spans the orthogonal complement to $L$, and $\displaystyle \| U \|^2 = \traceop\big[UU^\trop\big]$ is the (squared) Frobenius matrix norm. While $\gcop^*(X \to Y)$ will not in general vary monotonically with $\gcop(X \to Y)$, simulations indicate strongly that subspaces $L$ which locally minimise $\gcop(X \to Y)$ will lie in regions of the Grassmannian with near-locally-minimal $\gcop^*(X \to Y)$. Since $\gcop^*(X \to Y)$, as a biquadratic function of $L,M$, is considerably less computationally expensive to calculate\footnote{Note that the sequence $C A^k K$, $k = 0,\ldots,r-1$, may be pre-calculated for a given ISS model.} than $\gcop(X \to Y)$, we have found that pre-optimising $\gcop^*(X \to Y)$ under the constraints $LL^\trop = 0, LM^\trop = 0, MM^\trop = 0$ leads to significantly accelerated optimisation of $\gcop(X \to Y)$, especially for highly multivariate state-space models, and/or models with large state-space dimension. The same techniques may be used to optimise $\gcop^*(X \to Y)$ as $\gcop(X \to Y)$.

For optimisation of $\gcop(X \to Y)$, it may also be more computationally efficient to use the spectral integral form \eqref{eq:ssgcint} rather than \eqref{eq:ssdd} with the DARE \eqref{eq:dddares}. Approximating the integral will involve choosing a frequency resolution $d\omega$ for numerical quadrature. We have found that a good heuristic choice is $d\omega \approx \log\rho/\log\eps$, where $\rho$ is the spectral radius of the process (\secref{sec:cglss}), and $\eps$ the machine floating-point epsilon\footnote{The \emph{autocorrelation} at lag $k$ of a VAR process decays $\propto\rho^k$, and we note that by the well-known Wiener–Khintchine theorem, the CPSD $S\big(e^{-i\omega}\big)$ which appears under the integral in \eqref{eq:ssgcint} is the discrete Fourier transform (DFT) of the autocorrelation sequence. Thus according to the heuristic, any further precision in the quadrature conferred by finer spectral resolution is likely to be consumed by floating-point (relative) rounding error.}. Quadrature, then, is likely to be computationally cheaper than solving the DARE \eqref{eq:dddares} provided that $\rho$ is not too close to $1$, and for fixed $\rho$ scales better with system size.

For numerical minimisation of dynamical dependence derived from empirical data via state-space or VAR modelling, a stopping criterion may be based on statistical inference (\secref{sec:statsinf}): iterated search may be terminated on failure to reject the appropriate null hypothesis of vanishing dynamical dependence at a predetermined significance level. As mentioned in \secref{sec:statsinf}, in lieu of a known sampling distribution for the state-space Granger causality estimator, this is only likely to be practicable for VAR modelling.

So far, we have had some success (with both the redundant $L$-parametrisation under the orthonormality constraint and the $1\!-\!1$ $M$-parametrisation with on-the-fly $(\Psi,\Pi)$ selection), with (i) stochastic gradient descent with annealed step size, and, more computationally efficiently (ii) with a $(1\!+\!1)$ \emph{evolution strategy} \citep[ES;][]{Rechenberg:1973,Schwefel:1995} -- both with multiple restarts to identify local sub-optima. The search space scales effectively quadratically with $n$; so far, we have been able to solve problems up to about $n \approx 20$ (for all $m$) in under than $24$ hours on a standard multi-core Xeon\textsuperscript{\texttrademark} workstation; with pre-optimisation, we may extend this to about $n \approx 100$, although the number of local sub-optima increases with $n$, necessitating more restarts. Parallel high-performance computing aids significantly, since restarts are independent and may be run concurrently. GPU computing should also improve efficiency, in particular if GPU-enabled DARE solvers are available.

\subsection{Simulation results} \label{sec:disim}

In this Section we demonstrate the discovery of dynamically-independent macroscopic variables and estimation of the emergence portrait for state-space systems with specified causal connectivity, using numerical optimisation (\secref{sec:macglssn}). In an empirical setting, a state-space (or VAR) model for stationary data could be estimated by standard methods and the same optimisation procedure followed.

\begin{figure}
	\begin{center}
	\begin{subfigure}[c]{0.35\textwidth}
		\centering
		\includegraphics[width=0.65\textwidth]{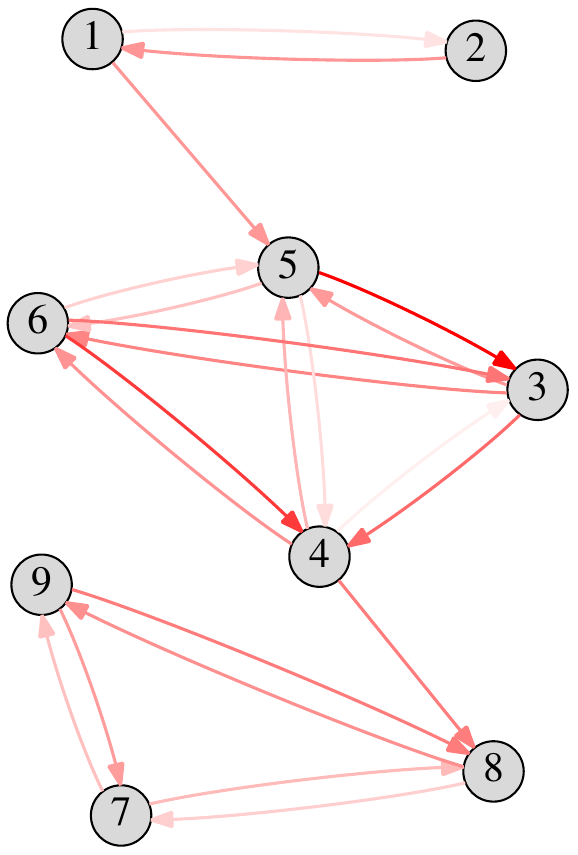}
		\caption{\footnotesize Granger-causal graph: arrow weight represents relative strength of the corresponding directed causality.} \label{fig:cgraph1a}
	\end{subfigure}
	\hfill
	\begin{subfigure}[c]{0.62\textwidth}
		\centering
		\parbox{\textwidth}{
		{\footnotesize \[
		\cgraphn = \begin{bmatrix*}[r]
			\mnul & 0.038 &     0 &     0 &     0 &     0 &     0 &     0 &     0 \\
			0.010 & \mnul &     0 &     0 &     0 &     0 &     0 &     0 &     0 \\
				0 &     0 & \mnul & 0.006 & 0.092 & 0.050 &     0 &     0 &     0 \\
				0 &     0 & 0.054 & \mnul & 0.012 & 0.071 &     0 &     0 &     0 \\
			0.037 &     0 & 0.036 & 0.027 & \mnul & 0.017 &     0 &     0 &     0 \\
				0 &     0 & 0.044 & 0.039 & 0.022 & \mnul &     0 &     0 &     0 \\
				0 &     0 &     0 &     0 &     0 &     0 & \mnul & 0.018 & 0.036 \\
				0 &     0 &     0 & 0.047 &     0 &     0 & 0.025 & \mnul & 0.047 \\
				0 &     0 &     0 &     0 &     0 &     0 & 0.023 & 0.040 & \mnul
		\end{bmatrix*}
		\]}}
		\vspace{3.2em}
		\caption{\footnotesize Pairwise-conditional Granger causality matrix: columns index source node, rows target node.} \label{fig:cgraph1b}
		\end{subfigure}
	\end{center}
	\vspace{-1.0em}
	\caption{\small Granger-causal structure for a $9$-variable VAR($7$) model comprising three fully-connected modules with two inter-module connections. The model was constructed by randomly generating autoregression coefficients $A_{k,ij}$, $k = 1,\ldots,7$, $i,j = 1,\ldots,9$, setting the coefficients to zero for ``missing'' connections (zeros in the matrix in \figref{fig:cgraph1b}), and normalising to spectral radius $\rho = 0.9$.} \label{fig:cgraph1}
\end{figure}

Our simulations are motivated as follows: if at scale $0 < m < n$ we have a macroscopic variable $Y_t = LX_t$, then (\cf~\secref{sec:cdi}) the system may be transformed so that the linear mapping $L$ takes the form of a \emph{projection} onto some $m$-dimensional coordinate hyperplane $x_{i_1},\ldots,x_{i_m}$, and according to \eqref{eq:dicg}, $Y_t$ is perfectly dynamically-independent iff
\begin{equation}
	\gcop(X \to Y) = 0 \iff \cgraphn_{ij} = 0 \quad\text{for}\quad i = i_1,\ldots,i_m\,, \forall j \label{eq:ddgccgx}
\end{equation}
where $\cgraphn$ is the causal graph \eqref{eq:tecgraph} of the system. While not an entirely general characterisation of the emergence portrait (as remarked at the end of \secref{sec:DI}, multiple dynamically-independent linear coarse-grainings will not in general be \emph{simultaneously} transformable to axes-hyperplane projections), we may nonetheless design linear models with prespecified causal graphs, which then mediate the expected dynamically-independent macroscopic variables at various scales. This construction is simple to achieve with finite-order VAR models (less so with more general state-space models\footnote{The condition \eqref{eq:ssdd0} for vanishing state-space GC with a given causal graph is highly nonlinear, and consequently difficult to enforce numerically.}), by setting the appropriate VAR coefficients $A_{k,ij}$ to zero. (We may also achieve ``near dynamical independence'' by making the appropriate $A_{k,ij}$ ``small''.) This is illustrated in \figref{fig:cgraph1}.
In an empirical setting, it would be preferable to ``prune'' the Granger-causal graph to only display statistically-significant pairwise-conditional Granger causalities as directed edges.

\begin{figure}
	\begin{center}
		\includegraphics{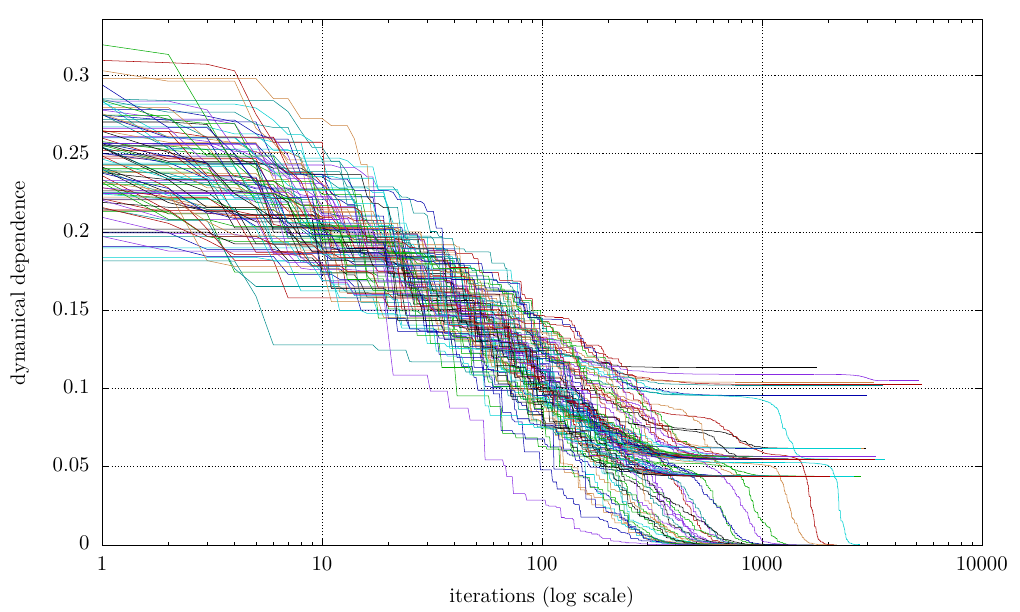}
	\end{center}
	\caption{\small Dynamical dependence minimisation for the $9$-variable VAR($7$) model in \figref{fig:cgraph1}: $100$ runs of the $(1\!+\!1)$-ES at scale $m = 6$. See main text for details.} \label{fig:opthist1}
\end{figure}

We minimised dynamical dependence at scales $m = 1,\ldots,8$ for the VAR model of \figref{fig:cgraph1}, using a $(1\!+\!1)$-ES (\secref{sec:macglssn}). At each scale, $100$ independent optimisation runs were performed, initialised uniformly randomly on the Grassmannian. The $(1\!+\!1)$-ES algorithm  was implemented as follows: initial step size is set to $\sigma = 0.1$. At each step, the current orthonormal $m \times n$ matrix $L$ representing the Grassmannian element is ``mutated'' by addition of a random $m \times n$ matrix $\Delta L$ with each element drawn independently from a $\mathcal N(0,\sigma^2)$ distribution. The mutant $L' = L+\Delta L$ is then othonormalised (using a singular value decomposition), and its dynamical dependence $d'$ calculated as in \secref{sec:diar}. If $d'$ is less than or equal to the current dynamical dependence $d$, then $L$ is replaced by $L'$, and the step size $\sigma$ increased by a multiplicative factor $\nu_+$; otherwise, the original $L$ is retained, and the step size decreased by a multiplicative factor $\nu_-$. The adaptation factors were calculated according to a version of the well-known Rechenburg ``$1/5$th success rule'' described in \citet[Sec.~3.1]{HansenEtal:2015}: a gain factor $\gamma$ is set to $1/\sqrt{\delta+1}$ where $\delta = m(n-m)$ is the dimension of the Grassmannian search space. We then set
\begin{subequations}
\begin{align}
		\nu_+ &= e^{(1-h) \gamma} \\
		\nu_- &= e^{-h \gamma};
\end{align} \label{eq:1in5}%
\end{subequations}
with $h = 1/5$. The algorithm is deemed to have converged\footnote{A time-out of $10,000$ iterations was set; in fact most runs converged within $1,000-3,000$ iterations, and $< 10$ failed to converge. The simulation, coded in MATLAB, required $\approx 40$ minutes computation time on a $12$-core Xeon workstation running Ubuntu Linux 16.04. For illustrative purposes, we did \emph{not} use the ``proxy cost function'' acceleration described in \secref{sec:macglssn}; experiments indicated that pre-optimisation speeds up optimisation roughly by a factor of $10$.} when either the step size $\sigma$ or the current dynamical dependence $d$ falls below a threshold value\footnote{In an empirical scenario, we might instead implement the inferential stopping criterion suggested in \secref{sec:macglssn}; that is, the threshold becomes the critical value for the sample Granger causality $d$ at a specified significance level.} set to $10^{-8}$. A typical set of $100$ runs at scale $m = 6$ is plotted in \figref{fig:opthist1}.
We see that while many runs converge to the true minimum---the unique zero-dynamical-dependence subspace (\cf~\figref{fig:localopt1} and \figref{fig:wgraph1g} below)---other runs become trapped in local sub-optima.

Full optimisation results across all scales are illustrated in \figref{fig:localopt1}, which may be considered as a graphical overview of the empirically-derived emergence portrait of the system in accordance with our \textit{ansatz} \eqref{txt:ansatz}:
\begin{figure}
	\begin{center}
		\includegraphics{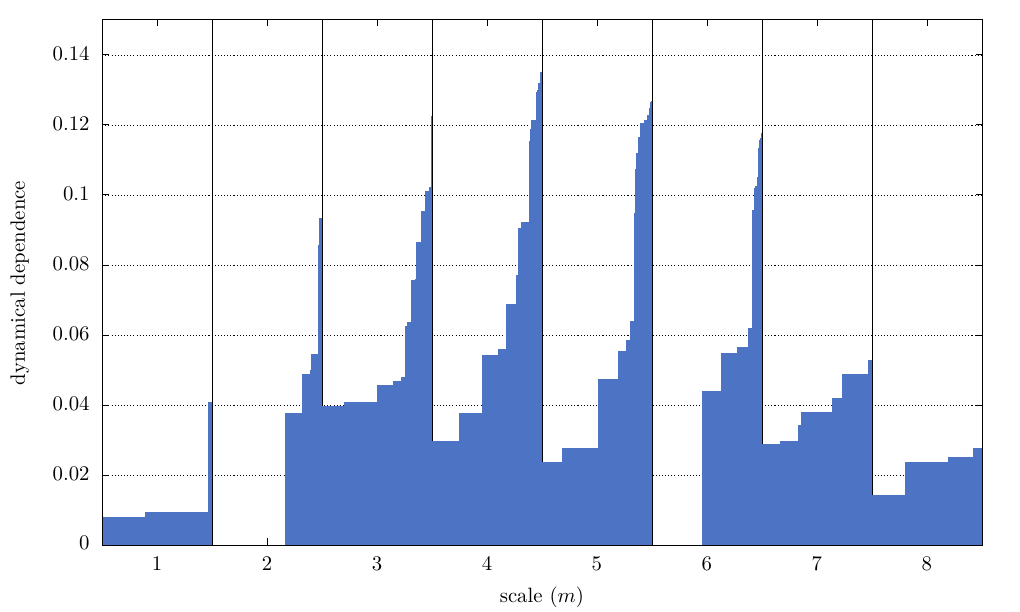}
	\end{center}
	\caption{\small Emergence portrait (I): results of $(1\!+\!1)$-ES dynamical dependence minimisation at all scales of the $9$-variable VAR($7$) model in \figref{fig:cgraph1}. At each scale, the heights of the bars indicate the sorted optimal (minimised) dynamical dependencies for $100$ runs with uniform random initialisation. See main text for details.} \label{fig:localopt1}
\end{figure}
at each scale the $100$ locally-optimal terminating values of dynamical dependence are sorted in ascending order and plotted on a bar chart. Zero values indicate the presence of perfectly dynamically-independent subspaces at the corresponding scale, while non-zero values indicate locally-optimal subspaces; near-zero values indicate ``nearly dynamically-independent'' projections. The width of ledges of equal dynamical dependence at each scale give an indication of the size of the basin of attraction of the local minimum.

We see clearly from the figure that, as expected from the causal graph (\figref{fig:cgraph1}) there are perfectly dynamically-independent macroscopic variables only at scales $2$ and $6$, corresponding to projection onto the sub-graphs $\{1,2\}$ and $\{1,2,3,4,5,6\}$ respectively (\cf~\figreff{fig:wgraph1a}{fig:wgraph1g} below); only these sub-graphs have no incoming Granger-causal connections from the rest of the graph. (It is important to note, though, that this ``no incoming connections'' for dynamically-independent subspaces holds for our simple example as a direct consequence of its construction from a causal graph; we are, in effect, working in a ``privileged'' coordinate system. For an arbitrary system, where we could not be expected to know \textit{a priori} which particular coordinate transformation(s) map the dynamically-independent subspaces to the causal graph as per eq.~\eqref{eq:ddgccgx}, this would no longer hold in general.)

\figref{fig:localopt1} on its own does not reveal the full detail of the emergence portrait; in particular, while indicating the broad distribution of dynamical dependence of (locally-)optimal subspaces (\ie, the macroscopic variables), it says little about the subspaces in relation to the system itself, or to each other -- for example, it is not clear whether bars of equal height at a given scale actually correspond to the \emph{same} subspace or not. To dig deeper into the emergence portrait, we need to consider the (locally-)maximally dynamically-independent subspaces explicitly in the structured domain in which they reside -- that is, the Grassmannian manifold of vector subspaces. Visualisation of vector subspaces in high-dimensional Euclidean spaces---elements of the Grassmannian---is challenging. Below we present a series of visualisations designed to aid intuition on the structure of, and relationships between, locally-optimal subspaces.

\begin{figure}
	\begin{center}
		\includegraphics{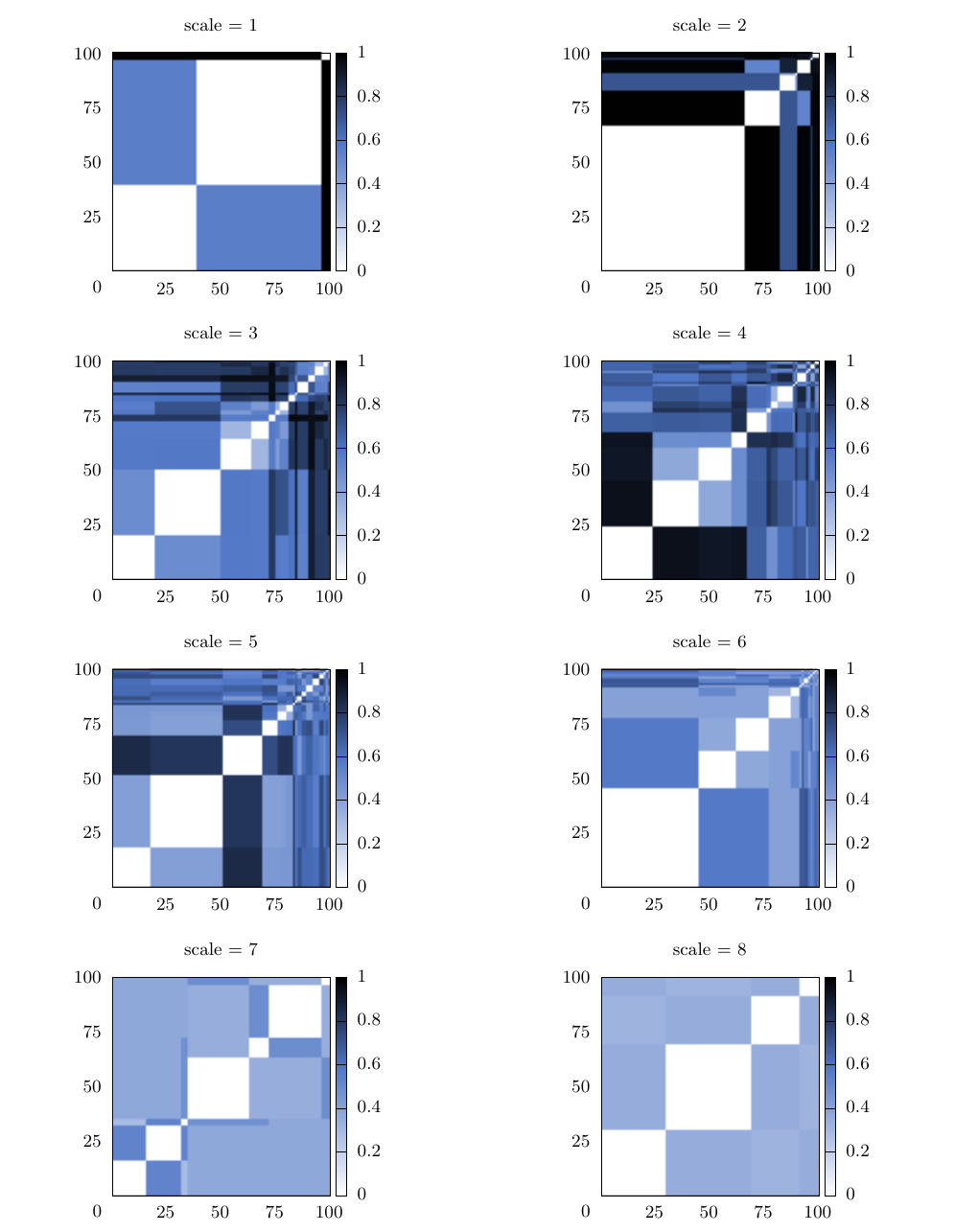}
	\end{center}
	\caption{\small Pair-wise inter-optimum distances between locally-optimal subspaces for the $9$-variable VAR($7$) model in \figref{fig:cgraph1}, for $100$ independent optimisation runs. See main text for details.} \label{fig:iodist1}
\end{figure}

Firstly, as alluded to in \secref{sec:cglss}, we may calculate the principal angles between two subspaces of $\reals^n$; in fact, for subspaces of dimensions $0 < m_1 \le m_2 < n$ there are $m_1$ principal angles $0 \le \theta_1 \le \ldots \le \theta_{m_1} \le \pi/2$ \citep{Wong:1967}, and we may define a metric on the Grassmannian---a measure the  distance between subspaces---as $\sqrt{\theta_1^2+\cdots+\theta_{m_1}^2}$, normalised by $\frac\pi 2\sqrt{m_1}$ to lie in $[0,1]$. We may use this metric to to answer the question posed above: do the locally-optimal subspaces of equal dynamical dependence at a given scale, as evidenced in \figref{fig:localopt1}, in general correspond to the same subspaces or not? To this end, at each scale we may calculate the distances between all pairs of the $100$ locally-minimal subspaces. In \figref{fig:iodist1} these distances are represented on a colour scale (this figure should be viewed alongside \figref{fig:localopt1}).
We see from the white squares on the diagonals that at each scale the local optima of equal dynamical dependence are in general zero distance from each other, and thus correspond to the \emph{same} local optimum. Further structural detail may be inferred from \figref{fig:iodist1}; for instance, from the $m = 2$ results we see that the second-lowest dynamical dependence subspace (around the $70-80$th run) is almost orthogonal to the zero-dynamical dependence subspace (\cf~\figreff{fig:wgraph1a}{fig:wgraph1b} below) -- that is, these subspaces are highly dissimilar.

To gain insight into the placement of locally-optimal subspaces with respect to the system itself, we may calculate the distances between an $m$-dimensional subspace and each of the coordinate axes $x_1,\ldots,x_n$ (in this case there is only a single principal angle). Note that these $n$ distances do not uniquely identify the Grassmannian element, unless $m = 1$ or $n-1$ (in higher dimensions there is more ``wiggle room'' for subspaces); however, if the distance between a subspace $L$ and a coordinate axis is zero, then we can conclude that that axis is co-linear with $L$. Thus by \eqref{eq:ddgccgx} it follows that  $Y_t = LX_t$ is perfectly dynamically-independent iff distances to some set of axes $x_{i_1},\ldots,x_{i_m}$ are zero ($L$ is then a projection onto the linear subspace spanned by those axes). Given a specified Granger-causal graph $\cgraphn$ and a linear subspace, we present it graphically as a weighted graph with edges coloured according to the pairwise-conditional Granger causalities, and nodes coloured according to the distance between the corresponding coordinate axis and the subspace. \figref{fig:wgraph1} displays some colour-weighted Granger-causal graphs corresponding to locally-optimal subspaces at scales $2$, $5$ and $6$.
\begin{figure}
	\def\figwidth {0.23\textwidth}
	\begin{center}
	\begin{subfigure}[b]{\figwidth}
		 \centering
		 \includegraphics[width=\textwidth]{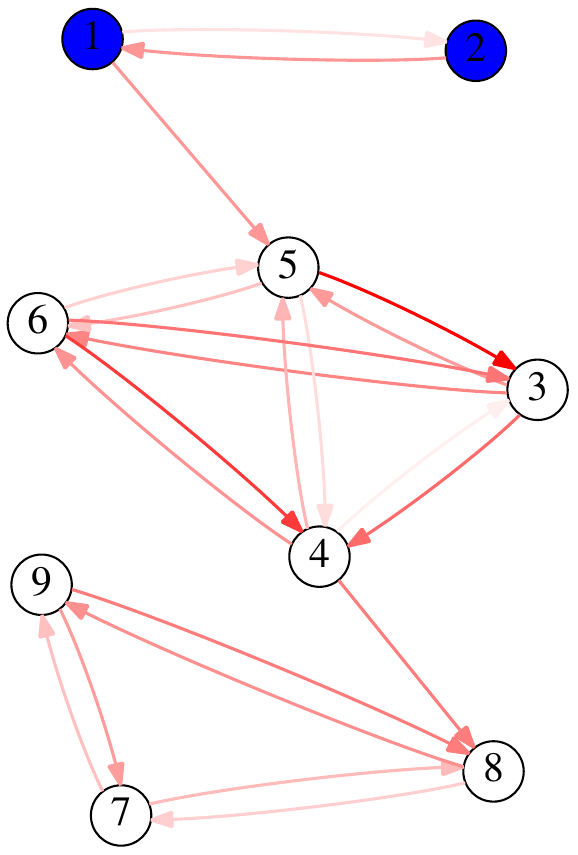}
		 \caption{scale $= 2$, run $1$} \label{fig:wgraph1a}
	\end{subfigure}
	\hfill
	\begin{subfigure}[b]{\figwidth}
		 \centering
		 \includegraphics[width=\textwidth]{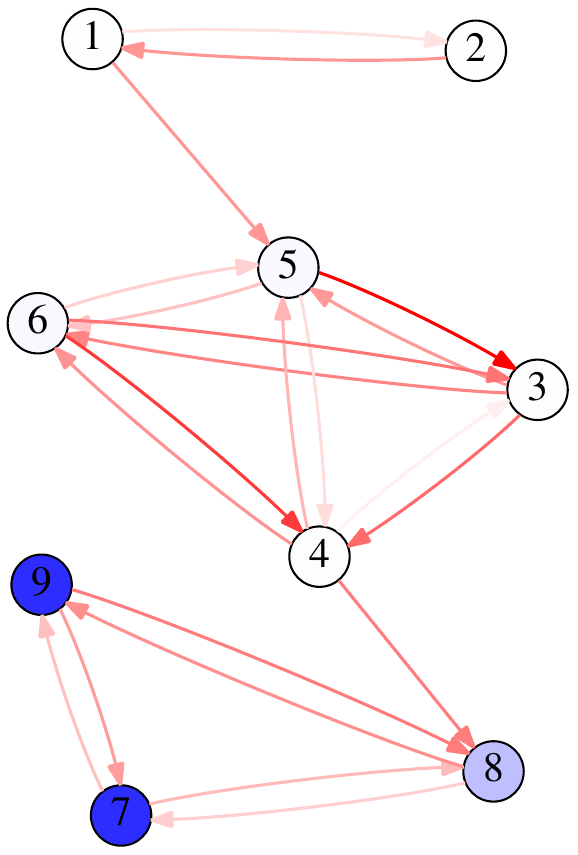}
		 \caption{scale $= 2$, run $70$} \label{fig:wgraph1b}
	\end{subfigure}
	\hfill
	\begin{subfigure}[b]{\figwidth}
		 \centering
		 \includegraphics[width=\textwidth]{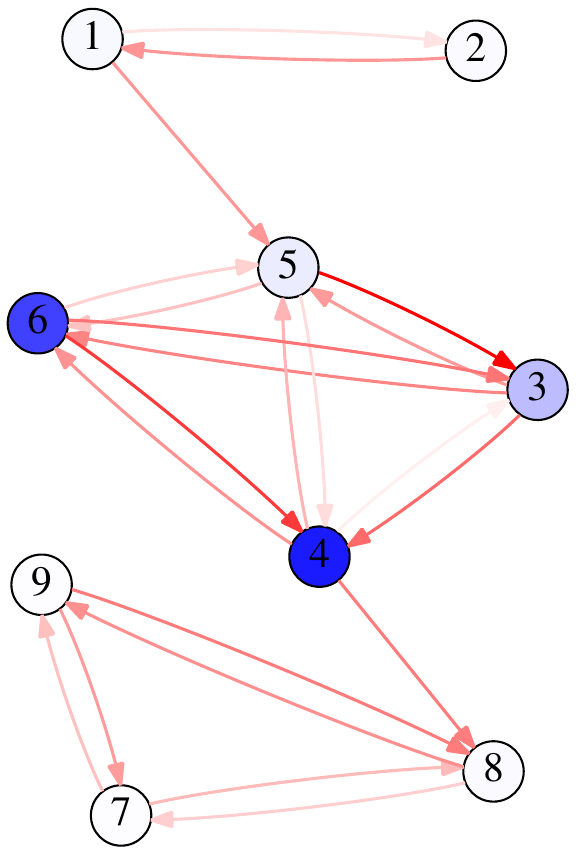}
		 \caption{scale $= 2$, run $99$} \label{fig:wgraph1c}
	\end{subfigure}
	\\[3em]
	\begin{subfigure}[b]{\figwidth}
		 \centering
		 \includegraphics[width=\textwidth]{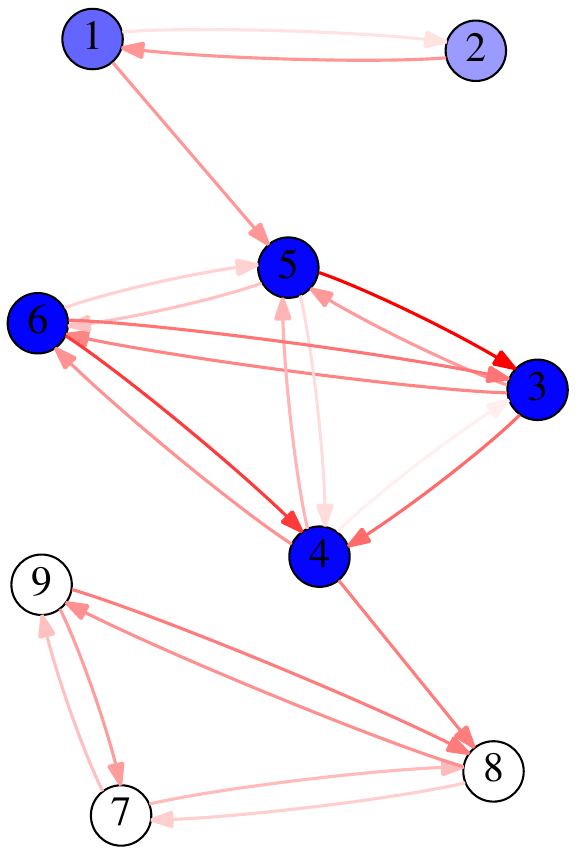}
		 \caption{scale $= 5$, run $1$} \label{fig:wgraph1d}
	\end{subfigure}
	\hfill
	\begin{subfigure}[b]{\figwidth}
		 \centering
		 \includegraphics[width=\textwidth]{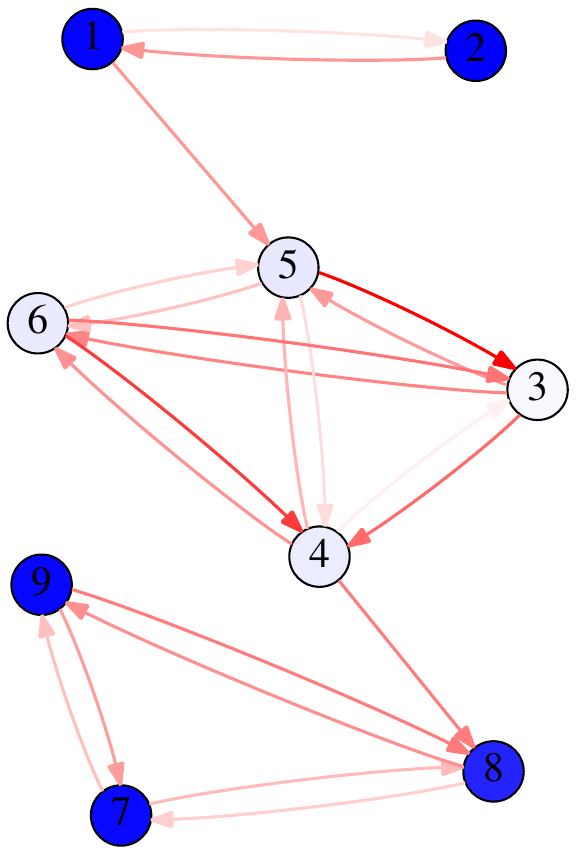}
		 \caption{scale $= 5$, run $60$} \label{fig:wgraph1e}
	\end{subfigure}
	\hfill
	\begin{subfigure}[b]{\figwidth}
		 \centering
		 \includegraphics[width=\textwidth]{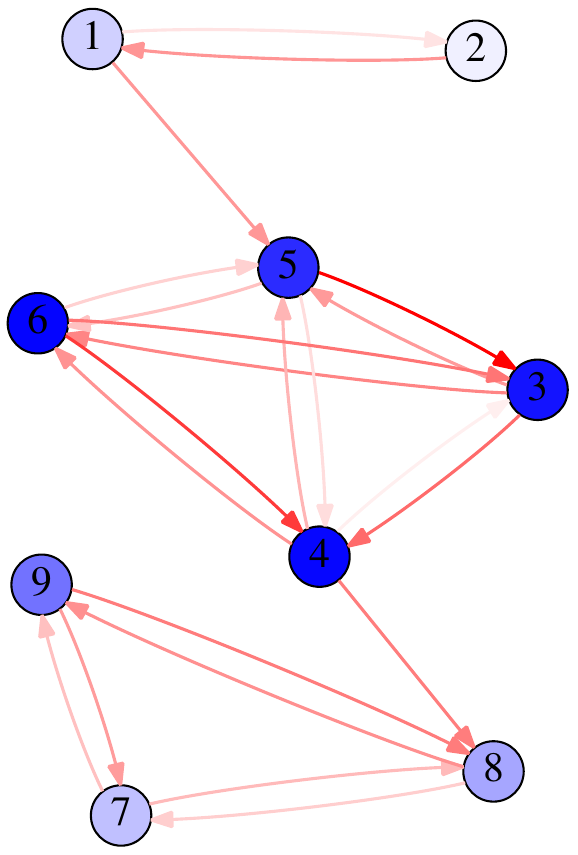}
		 \caption{scale $= 5$, run $80$} \label{fig:wgraph1f}
	\end{subfigure}
	\\[3em]
	\begin{subfigure}[b]{\figwidth}
		 \centering
		 \includegraphics[width=\textwidth]{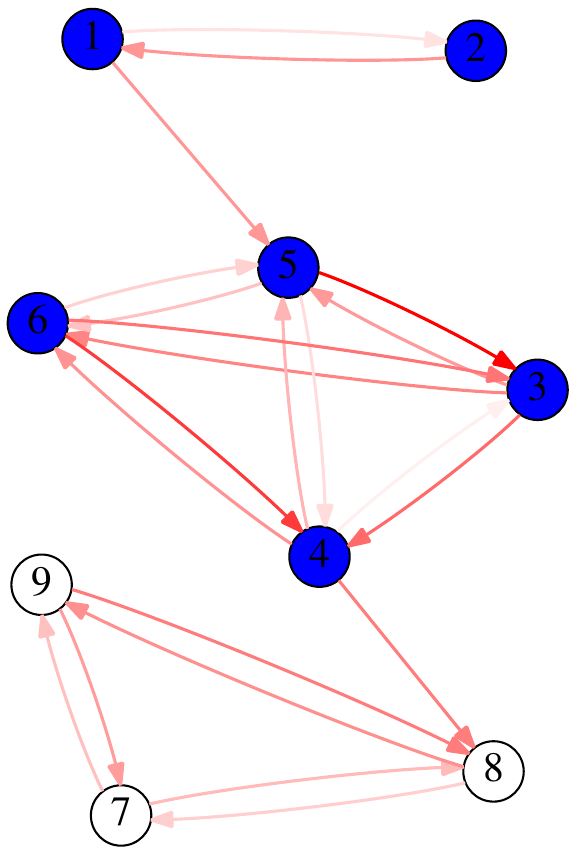}
		 \caption{scale $= 6$, run $1$} \label{fig:wgraph1g}
	\end{subfigure}
	\hfill
	\begin{subfigure}[b]{\figwidth}
		 \centering
		 \includegraphics[width=\textwidth]{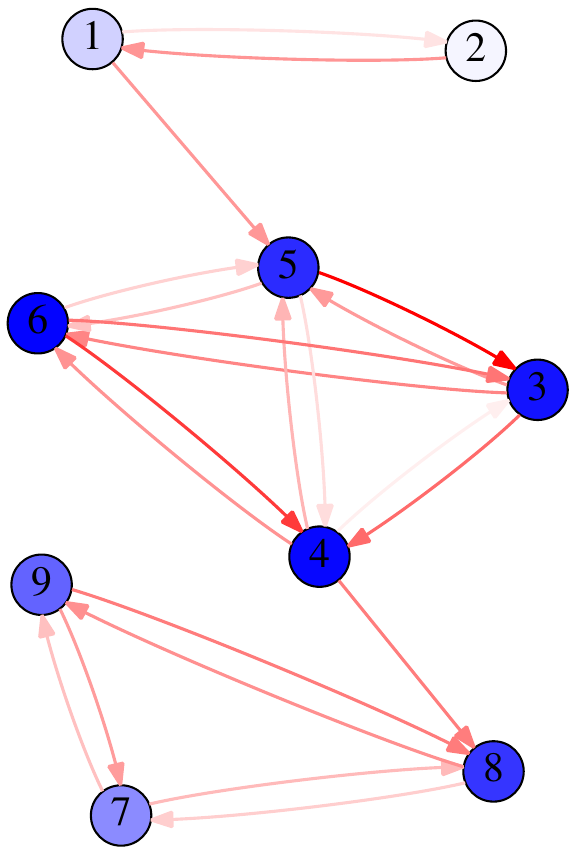}
		 \caption{scale $= 6$, run $50$} \label{fig:wgraph1h}
	\end{subfigure}
	\hfill
	\begin{subfigure}[b]{\figwidth}
		 \centering
		 \includegraphics[width=\textwidth]{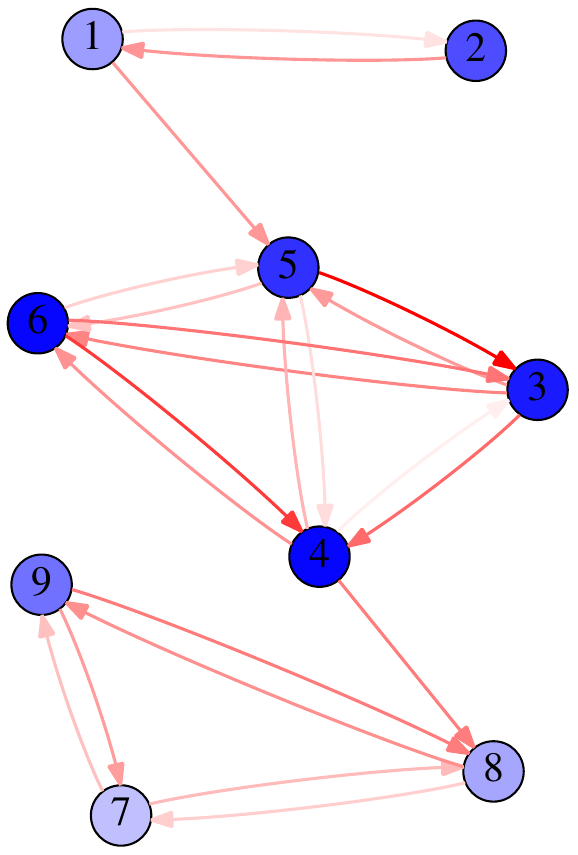}
		 \caption{scale $= 6$, run $80$} \label{fig:wgraph1i}
	\end{subfigure}
	\\[1em]
	\caption{\small Granger-causal graphs for locally-optimal subspaces with nodes colour-weighted by axis angle, for $9$-variable VAR($7$) model in \figref{fig:cgraph1}. A solid blue node indicates that the subspace is co-linear with the corresponding axis, while a white node indicates that it is orthogonal to that axis. Note that, since runs are sorted in ascending order of dynamical dependence, higher run numbers correspond to \emph{less} dynamically-independent subspaces. See main text for details.} \label{fig:wgraph1}
	\end{center}
\end{figure}
A solid blue node indicates that the subspace is co-linear with the corresponding axis, a white node that it is orthogonal to that axis, while intermediate shades indicate angles between zero and $\pi/2$.

The top row in \figref{fig:wgraph1} displays weighted graphs for $2$-dimensional subspaces discovered by the ES, with locally-minimum dynamical dependence. \figref{fig:wgraph1a} indicates the unique $2$-node sub-graph corresponding to a perfectly dynamically-independent $2$-dimensional subspace. \figref{fig:wgraph1b} illustrates a nearly-dynamically-independent $2$-dimensional subspace; the subspace corresponding to \figref{fig:wgraph1c}, while a local minimum, is not very dynamically-independent. The middle row of figures display graphs for $5$-dimensional subspaces; there are no perfectly dynamically-independent subspaces at this scale, but the subspace corresponding to \figref{fig:wgraph1d} is, again, nearly dynamically-independent, the subspaces corresponding to the graphs to its right less so. \figref{fig:wgraph1g} identifies the unique perfectly dynamically-independent subspace at scale $6$, the graphs to its right local dynamical dependence sub-optima.

\begin{figure}
	\begin{center}
		\includegraphics{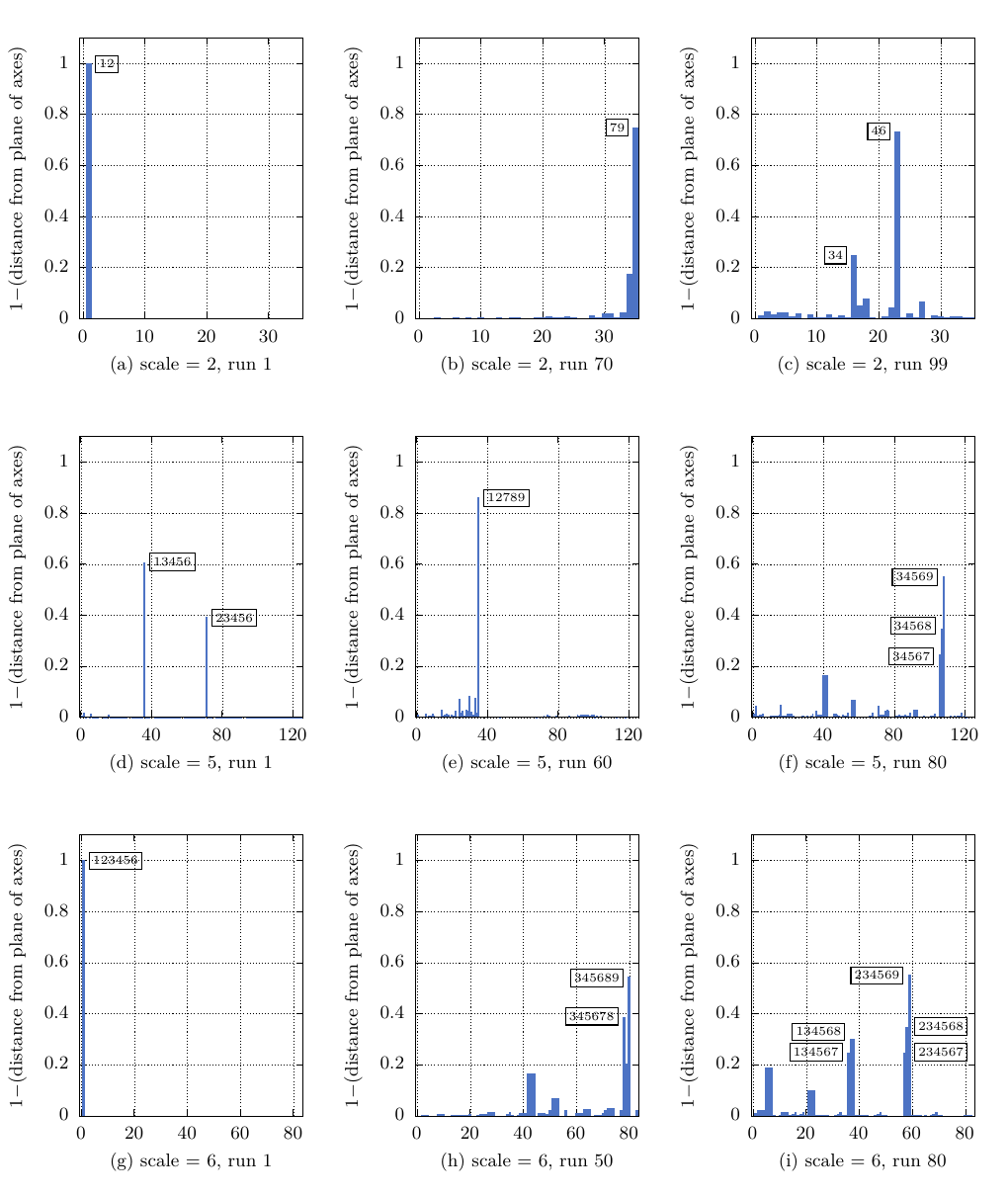}
	\end{center}
	\caption{\small Co-planarity of locally-optimal subspaces with subspaces of the same dimension spanned by coordinate axes, for the $9$-variable VAR($7$) model in \figref{fig:cgraph1}. The height of the bars is $1 - \theta_{\text{max}}$, where $0 \le \theta_{\text{max}} \le 1$ is the normalised maximum principal angle between the locally optimum subspace and corresponding coordinate subspace; so $1$ indicates co-planarity, $0$ orthogonality.  The horizontal scale is the dictionary ordering of the  $\binom n m$ axis combinations. Boxed labels show x-axis numbers of bars; \eg, the ``spike'' in subfigure (e) at ordinal $35$ corresponds to the subspace spanned by the axes $x_1, x_2, x_7, x_8, x_9$ (\cf~\figref{fig:wgraph1e}). See main text for details.} \label{fig:plucker1}
\end{figure}

The previous visualisation examined subspace (angular) distance from coordinate axes. This still allowed for a lot of ``wiggle room'': at least in higher dimensions, many subspaces of a given dimension are equidistant from all the coordinate axes. In the next, finer-grained visualisation, we look instead at subspace distance from coordinate \emph{hyperplanes} of the same dimension as the subspace, \ie, subspaces spanned by subsets of the coordinate axes. Now there is no ``wiggle room''; a subspace is uniquely identified by its distances from all same-dimension coordinate hyperplanes. In \figref{fig:plucker1}, for the same examples as in \figref{fig:wgraph1}, at the appropriate scale $m$ we plot a measure of how co-planar the locally-optimal subspace is with each of the $\binom n m$ subspaces spanned by combinations $x_{i_1},\ldots,x_{i_m}$, $1 \le i_1 < \ldots < i_m \le n$, of the coordinate axes. The horizontal scale is the dictionary (lexicographic) ordering of the axis combinations. The height of the bars is $1 - \theta_{\text{max}}$, where $0 \le \theta_{\text{max}} \le 1$ is the normalised maximum principal angle between the locally optimum subspace and corresponding coordinate subspace\footnote{Here we use the \emph{maximum} principal angle rather than the root sum-of-squares metric, as (somewhat counter-intuitively), in higher dimensions two orthogonal $m$-dimensional subspaces may have some non-zero principal angles.}; thus $1$ indicates that the subspaces are co-planar, $0$ that they are orthogonal.
The figure may be compared with \figref{fig:wgraph1}, though here, by contrast, the metrics for each plot collectively identify a unique Grassmannian element. Taking some examples, for the locally-minimum $5$-dimensional subspace represented in in \figref{fig:wgraph1d}, we find that the two large bars in \figref{fig:plucker1}d correspond to the coordinate subspaces $x_1, x_3, x_4, x_5, x_6$ and $x_2, x_3, x_4, x_5, x_6$, which is not apparent from \figref{fig:wgraph1d} alone. The single high bar in \figref{fig:plucker1}e corresponds to the subspace spanned by the axes $x_1, x_2, x_7, x_8, x_9$---this is perhaps more apparent from \figref{fig:wgraph1e}---while the subspace in Figs.~\ref{fig:wgraph1f}/\ref{fig:plucker1}f exhibits a more complex relationship to the coordinate subspaces, with the highest co-planarity at $x_3, x_4, x_5, x_6, x_9$. Comparing Figs.~\ref{fig:plucker1}a and \ref{fig:plucker1}g confirms that the unique scale 2 dynamically-independent variable is nested in the unique scale 6 variable.

To summarise our analysis of the emergence portrait of the VAR($7$) system: \figref{fig:localopt1} displays just the distribution of dynamical dependence values of locally-optimal subspaces at each scale, as discovered by independent optimisation runs; in \figref{fig:iodist1} we examine at each scale the distances between discovered locally-optimal subspaces, thus enabling us to distinguish which represent \emph{unique} subspaces; in \figref{fig:wgraph1}, we measure the distances between optimal subspaces and coordinate axes, allowing a rough graphical depiction of their relationship to nodes on the causal graph; \figref{fig:plucker1} takes this a step further, pinpointing the exact positioning of the locally-optimal subspaces with respect to the coordinates of the microscopic system space.

Although avowedly a low-dimensional toy model, our analysis of the VAR($7$) system of \figref{fig:cgraph1} presents a viable approach to discovery of macroscopic variables in linear systems, and in accordance with our \textit{ansatz} \eqref{txt:ansatz}, grants insight into the local dynamical independence structure of linear systems. Our analysis also illustrates a more general point: developing a comprehensive \emph{emergence portrait}---how dynamically-independent macroscopic variables relate to the microscopic system and to each other---involves exposing the structure of the space of macroscopic variables. Thus in Figs.~ \ref{fig:localopt1}, \ref{fig:iodist1}, \ref{fig:wgraph1} and \ref{fig:plucker1}, we ``drill down'' into this structured space using progressively more detailed metrics. For real-world systems, relating the emergence portrait to macroscopic phenomenology is likely to be an empirical question -- but here too, understanding the structure of the emergence portrait \textit{in abstracto} is likely to be crucial.

\section{Deterministic and continuous-time dynamics} \label{sec:detcon}

Although the main thrust of this article concerns dynamical independence for discrete-time stochastic systems (\secref{sec:DI}), and in particular discrete-time linear systems (\secref{sec:cglss}), many systems commonly associated with emergent phenomena feature deterministic and/or continuous-time dynamics. For deterministic systems the question immediately arises as to how the information-theoretic framework of \secref{sec:DI} might apply, since Shannon information is indeterminate  for non-stochastic variables. In continuous time, furthermore, transfer entropy is more nuanced \citep{SpinneyProkopenkoLizier:2017} and considerably more complex, even in the linear case \citep{Barnett:downsample:2017}. Below we preview our approach to these important challenges.

\subsection{Deterministic dynamics in discrete time} \label{sec:detdisc}

For many discrete-time dynamical systems of interest, such as cellular automata, flocking models and discrete-time chaotic systems, dynamics take the deterministic Markovian form
\begin{equation}
	x_{t+1} = \xi(x_t) \label{eq:ddtn}
\end{equation}
with state transition function $\xi : \xstate \to \xstate$, where the microscopic state space $\xstate$ may be continuous and measurable, or discrete and countable. Thus given some initial condition $x_0$ we have $x_t = \xi^t(x_0)$, $t \ge 0$ where $\xi^t$ denotes $t$ iterations of the mapping $\xi$.

For discrete-state systems, we may consider the dynamics \eqref{eq:ddtn} with \emph{stochastic initial conditions}. Thus a random variable $X_0$ on $\xstate$ is introduced to represent the statistical distribution of initial ($t = 0$) microscopic states, yielding the microscopic stochastic process $X_t = \xi^t(X_0)$, $t \ge 0$, on $\xstate$. Given a coarse-graining $f : \xstate \to \ystate$, dynamical independence for the macroscopic variable $Y_t = f(X_t)$ may then be analysed along the lines of the general discrete-time stochastic case (\secref{sec:DI}). In practice, choice of the initial distribution may be based on general principles (\eg, maximum-entropy), or on domain-specific \textit{a priori} considerations. Then, since $Y_t$ depends deterministically on $X_t^- = \{X_0,X_1,\ldots,X_{t-1}\}$, the dynamical dependence \eqref{eq:tedd} of $Y$ on $X$ at time $t \ge 0$ is given simply by
\begin{equation}
	\teop_t(X \to Y) = \enop(Y_t | Y_t^-) = \enop(Y_{t+1}^-)-\enop(Y_t^-) \label{eq:ddtdd}
\end{equation}
Given the probability distribution (or density) function $p(x_0)$ for $X_0$, a general expression for the entropy $\enop(Y_{t+1}^-) = \enop(Y_0,Y_1,\ldots,Y_t)$, and thence $\teop_t(X \to Y)$, may be calculated.

This approach, though, is unviable for continuous-state systems of the form \eqref{eq:ddtn}, since for deterministic dynamics the transfer entropy $\teop_t(X \to Y)$ based on \emph{differential} entropies diverges\footnote{If $X$ is a random variable on $\reals^n$ and $f : \reals^n \to\reals^m$ with $m < n$, then unlike the discrete-state case, the differential conditional entropy $\enop(f(X)|X)$ is not zero, but rather diverges to $-\infty$.}. An alternative approach is to introduce \emph{scalable noise} into the process \eqref{eq:ddtn}, and then analyse dynamical independence for the resultant discrete-time stochastic system in the limit of vanishing noise. If the state space $\xstate$ is Euclidean, for example, we might consider the autoregressive stochastic process $X_{t+1} = \xi(X_t) + \sigma\beps_t$ derived from \eqref{eq:ddtn}, in the limit $\sigma \to 0$, where $\beps_t$ is a multivariate-normal white noise.

\subsection{Flows: deterministic dynamics in continuous time} \label{sec:flow}

An important case of deterministic continuous-time dynamics is that of \emph{flows}, defined as the group action $\xi : \xstate \times \reals \to \xstate$ of the additive group $\reals$ on a set $\xstate$, such that
\begin{subequations}
\begin{align}
	\xi(\bx,0) &= \bx \label{eq:flowa} \\
	\xi\big(\xi(\bx,s),t\big) &= \xi(\bx,s+t) \label{eq:flowb}
\end{align} \label{eq:flow}%
\end{subequations}
for $\bx \in \xstate$, $s,t \in \reals$. If $\xstate$ is a differentiable manifold then the flow is \emph{smooth} if the function $\xi$ is differentiable, and for any fixed $t$ the function $\bx \mapsto \xi(\bx,t)$ is a diffeomorphism. If $\bx \mapsto \xi(\bx,t)$ is only a diffeomorphism on a strict subset of $\xstate \times \reals$, then $\xi$ is said to define a \emph{local} flow; from now on, we use the term ``flow'' to include local flows. On Euclidean space $\xstate = \reals^n$, smooth flows are essentially equivalent to $1$st-order autonomous ordinary differential equations (ODEs); the \emph{trajectory} $\bx(t) = \xi(\bx_0,t)$ is the unique solution of the autonomous ODE\footnote{A dot indicates [partial] differentiation with respect to $t$.} $\dot\bx(t) = g(\bx)$ with initial condition $\bx(0) = \bx_0$, where $g(\bx) = \dot\xi(\bx,0)$. Many classical dynamical systems, such as Hamiltonian mechanics, flocking, and chaotic dynamical systems, are expressed as ODEs and may thus be considered as flows.

For flows, stochastic initial conditions (\secref{sec:detdisc}) once again run up against problems with diverging differential entropies. As a potential remedy, we reconsider the original expression \eqref{eq:dians1} of dynamical independence. There, independence is interpreted in a statistical sense; here we propose a ``functional'' interpretation more aligned with dynamical systems theory: given a smooth flow $\xi : \reals^n \times \reals \to \reals^n$ at the microscopic level, a differentiable coarse-graining function $f : \reals^n \to \reals^m$, $0 < m < n$, is considered to define a dynamically-independent macroscopic variable for $\xi$ iff there is a flow $\eta : \reals^m \times \reals \to \reals^m$ on the macroscopic space such that\footnote{This is again comparable in spirit to \citet{AllefeldAtmanspacherWackermann:2009}, where the parsimony of macroscopic variables is associated with the preservation of a Markov property.}
\begin{equation}
	 f\big(\xi(\bx,t)\big) = \eta\big(f(\bx),t\big) \label{eq:ctdi}
\end{equation}
for all $\bx \in \reals^n$ and $t \in \reals$. In terms of ODEs, this is equivalent to the existence of an autonomous ODE\footnote{Note that ``autonomous'' with reference to an ODE indicates that the equation has no explicit time dependence.} $\dot\by(t) = h(\by) \label{eq:yde1}$ on $\reals^m$ such that for any trajectory $\bx(t)$ of $\xi$, $\by(t) = f\big(\bx(t)\big)$ is a trajectory of $\eta$. Thus, in the spirit of \eqref{eq:dians1}, a dynamically-independent macroscopic system is self-determining: given an initial condition $\by_0 \in \reals^m$, the coarse-grained macroscopic system determines its evolution in time without reference to the micro-level dynamics. Dynamical independence in this sense is invariant with respect to smooth coordinate transformations of both the microscopic space $\reals^n$ and the macroscopic space $\reals^m$; dynamical independence may thus be extended to flows on \emph{differentiable manifolds} via overlapping coordinate charts \citep{KobayashiNomizu:1996}.

In preliminary work (in preparation) we derive necessary and sufficient condition for dynamical independence in the above sense, and show that  dynamically-independent coarse-grainings $f : \reals^n \to \reals^m$ are built from \emph{invariants} (conserved quantities)  of the flow, along with a ``time-like'' scalar function; see \apxref{sec:hamdyn} for a summary of these results. Thus dynamical independence in the functional sense essentially reduces to the classical problem of invariants of flows on differentiable manifolds \citep{Cohen:1911}; \cf~the example of a Newtonian galaxy at the beginning of \secref{sec:DI}. By the celebrated First Theorem of Noether \citep{Arnold:1978}, for Lagrangian systems invariants are associated with symmetries of the Lagrangian action. Thus for such systems Noether's Theorem characterises the dynamically-independent macroscopic variables. However, by no means all systems of  interest fall into this class. (The central role of symmetry in Noether's Theorem, though, seems worth bearing in mind.)

The functional approach has one drawback: unlike the transfer entropy measure \eqref{eq:tedd} in the discrete-time stochastic case, it lacks an information-theoretic interpretation, and does not yield up an obvious candidate measure for dynamical dependence, let alone a  transformation-invariant one (we are currently investigating whether such a measure may exist); there is thus no ready notion of ``near-dynamical independence''. As for the discrete-time case (\secref{sec:detdisc}), there is also the possibility of adding scalable noise to the ODE to derive a \emph{stochastic differential equation} \citep[SDE;][]{Oksendal:2003}, and then consider the limiting behaviour of the micro $\to$ macro transfer entropy measure in the limit of vanishing noise. Nonlinear SDEs, though, are challenging to analyse in any generality.

\section{Discussion} \label{sec:discussion}

In this paper we introduce a notion of emergence of macroscopic dynamical structure in highly-multivariate microscopic dynamical systems, which we term \emph{dynamical independence}. In contrast to other treatments of emergence, which are largely concerned with part-whole and synergistic relationships between system components (see the discussion below), dynamical independence instantiates the intuition of an emergent process at a macroscopic scale as one which evolves over time according to its own dynamical laws, distinct from and independently of the dynamical laws operating at the microscopic level. More specifically, while prescribed by the microscopic process, a dynamically-independent macroscopic process is, conditional on its own history, independent of the history of the microscopic process. Dynamical independence is quantified by a Shannon information-based (and hence transformation-invariant) measure of dynamical dependence. Importantly, dynamical independence may be conditional on a co-distributed externally-demarcated process, thus accommodating systems which feature input-output interactions with a dynamic environment.

Critical to any theory of emergence over a potential range of spatiotemporal scales, is how we should construe a ``macroscopic variable''. Here, we try to keep this question as open as possible, with one key constraint: a macroscopic variable may be any process co-distributed with the microscopic process which, in predictive terms, does not ``bring anything new to the table'' beyond the microscopic: a macroscopic variable is prescribed by the microscopic process in the sense that it does not self-predict beyond the prediction afforded by the microscopic variables (along with the environment). We might thus conclude that if a macroscopic process appears to emerge as a process in its own right---with a ``life of its own''---this apparent autonomy is in the eye of a beholder blind to the micro-level dynamics. Emergence, in our approach, is therefore best thought of as being associated with particular ``ways of looking'' at a system.

A key aspect of our approach is an emphasis on \emph{discovery} of emergent macroscopic variables---``ways of looking'' at the system---given a micro-level description. Although specific problem domains may present ``natural'' prospective emergent macroscopic variables (which may be tested for degree of emergence by our dynamical dependence measure), this is by no means always the case. For neural systems, for example, it is in general far from clear how to identify candidate emergent processes. Groups of neurons firing in synchrony might intuitively suggest an emergent variable, but there may be many more (and more subtle) patterns of neural activity that may count as emergent without being intuitively apparent to an observer.  Our approach addresses this issue by ``automating'' the discovery process, through consideration of the full space of all admissible macroscopic variables; discovery of emergent variables then becomes a search/optimisation problem across this space. We introduce an \textit{ansatz} that proposes the results of this search, across all scales, as an informative account of the emergence structure of the given system - an ``emergence portrait''. Parametric modelling, furthermore, opens up the possibility of \emph{data-driven} discovery of emergent variables. We present an explicit calculation of dynamical dependence, and a detailed account of the search/optimisation process, for the important class of linear state-space models, suitable for wide deployment across a range of domains, including neural systems.

\subsection{Related approaches} \label{sec:relapp}

One difference between our approach and many related approaches concerns the role of the environment, and in particular the system/environment distinction \citep{KrakauerEtal:2020}.  Another is our emphasis on discovery of emergent phenomena, whereas the majority of approaches, while furnishing criteria or metrics for emergence, do not specify how, for a given microscopic process, candidate emergent processes might actually be found in practice.

Our notion of dynamical independence bears a resemblance to \emph{informational closure}, introduced by \citet{BertschingerEtal:2006}; a process $Y$ is described as ``informationally closed'' with respect to an environment $\envi$ when the transfer entropy $\teop(\envi \to Y)$ vanishes; that is, $Y$ is dynamically-independent with respect to the environment. To compensate for ``trivial'' systems where environment $\envi$ and system $Y$ are independent, this quantity is then subtracted from the mutual information $\miop(Y_t : E_t^-)$ to yield the ``non-trivial informational closure'' (NTIC), which may also be expressed as $\miop(Y_t : Y_t^-) - \miop(Y_t : Y_t^- \vbar E_t^-)$.

\citet{ChangEtal:2020} apply NTIC specifically to the case of coarse-grained macroscopic variables in the context of an environment. Their definition of a \emph{C-process} requires that the macroscopic variable $Y$ be (i) dynamically-independent of the system-environment ``universe'' $(X,E)$, and (ii) NTIC with respect to the environment $E$. Note that condition (i) is not equivalent to dynamical independence of $Y$ with respect to the system in the context of the environment.  While \citet{ChangEtal:2020} associate a C-process with a measure of consciousness, it is perhaps more generally (and less contentiously) construed as a notion of autonomy or emergence in complex systems.

Another relevant construct is \emph{G-emergence} \citep[Granger emergence;][]{Seth:2010}. \citet{Seth:2010} firstly operationalises the ``self-causation'' or ``self-determination'' of a variable $Y$ with respect to an external (multivariate) variable $Z$ as \emph{G-autonomy}\footnote{\citet{Seth:2010} expresses G-autonomy and G-emergence in terms of linear prediction; here we present the information-theoretic analogues under Gaussian assumptions; \cf~\secref{sec:cglss} below. In Seth's 2010 approach, the linear (Granger causality) formulation is important because his measure relies on a systematic \emph{in}ability to capture the full dynamical behaviour of a target system.}
\begin{equation}
	ga(Y \vbar Z) = \miop(Y_t : Y_t^- \vbar Z_t^-) = \enop(Y_t \vbar Z_t^-) - \enop(Y_t \vbar Z_t^-,Y_t^-) \label{eq:gautonomy}
\end{equation}
which measures the the degree to which inclusion of its \emph{own} past enhances prediction of $Y_t$ by the past of the external variable $Z_t$. Given a microscopic process $X_t$ and a macroscopic process $Y_t$, the G-emergence of $Y$ from $X$ is then specified as\footnote{Again, we have translated this into equivalent information-theoretic terms.}
\begin{equation}
	ge(Y \vbar X) = ga(Y \vbar X) + \teop(X \to Y) \label{eq:gemergence}
\end{equation}
This expression operationalises the notion that an emergent macroscopic process is, in a predictive sense, at once autonomous from, but also dependent on, the microscopic process -- again recalling the conceptual definition of ``weak emergence'' from \citep{Bedau:1997}. G-emergence differs from dynamical independence in two main respects. First, it requires that macroscopic variable be non-trivially self-predictive. Second, it includes a micro-to-macro term to assure, in an ad-hoc way, that they are related, in contrast to the principled approach to coarse-graining taken by dynamical independence.

We recognise immediately the second (transfer entropy) term in \eqref{eq:gemergence}---designed to ensure that the macro and the micro are related---as our dynamical dependence \eqref{eq:tedd} in the absence of a coupled environment, although for G-emergence it ``pulls in the opposite direction'', in the sense that increasing $\teop(X \to Y)$ increases G-emergence, but \emph{de}creases dynamical independence. Note also, though, that our requirement \eqref{eq:macrovar} on macroscopic variables---which holds in particular for coarse-grained variables---actually stipulates (in the absence of an environment) that the G-autonomy contribution $ga(Y \vbar X)$ in \eqref{eq:gemergence} vanishes identically, thus leaving G-emergence as precisely our dynamical \emph{dependence} rather than \emph{in}dependence, for the situations we consider for dynamical independence.

A recent approach with both parallels and differences to ours is that of \citet{RosasEtal:2020}. In contrast to our approach, their concern is explicitly with mereological (part-whole) causal relationships, such as \emph{downward causation}, what they term \emph{causal decoupling} and, in particular, \emph{causal emergence}. The latter is quantified as the unique predictive capacity of a supervenient feature over the microscopic system, beyond the predictive capacity of (parts of) the microscopic system. This is almost the obverse of dynamical independence, which hinges on prediction  of the \emph{macroscopic} rather than the microscopic process. Supervenience for ``features'' as defined by \citet{RosasEtal:2020}, it should be noted, does not generally correspond to our notion of supervenience for macroscopic variables. In contrast to our supervenience condition \eqref{eq:macrovar}, the comparable condition in \citep[Sec.~II]{RosasEtal:2020} is, in our notation
\begin{equation}
	\miop(X_t \ibar Y_t^- \vbar X_t^-) \equiv 0 \label{eq:superv}
\end{equation}
Although coarse-grained variables trivially satisfy both \eqref{eq:macrovar} and \eqref{eq:superv}, the latter again speaks to prediction of the microscopic, rather than macroscopic variable.

In order to express causal emergence in information-theoretic terms, \citet{RosasEtal:2020} make use of a \emph{partial information decomposition} \citep[PID;][]{WilliamsBeer:2010,WibralEtal:2017}. One challenge for this approach is a lack of consensus on what a ``canonical'' PID might look like. Further, current PID candidates tend to be computationally intractable and scale poorly with system size and macroscopic scale. In addition, the proposed measures are frequently framed in terms of discrete-valued (often finite) systems, and it is often unclear how they might be realised---or they become counter-intuitive and/or exhibit discontinuous behaviour---when extended to continuous-valued variables \citep{Barrett:2015}. Connected with the last point, many (though not all) lack the transformation invariance of Shannon information \citep{ChicharroEtal:2018, RosasEtal:2020a}. In recognition of the computational burden attached to PIDs, \citet{RosasEtal:2020} define Shannon information-based ``large system approximations'' for their measures, although it is unclear to what extent these reflect the intent of the respective PID formulations.

Closer in spirit to our approach is the theory of emergent brain macrostates propounded by \citet{AllefeldAtmanspacherWackermann:2009}. Along similar lines to dynamical independence, they consider dynamics for macroscopic systems which are in a sense ``self-contained'' with respect to the microscopic dynamics; however, unlike our more general information-theoretic approach, they associate such dynamics with a ($1$st-order, discrete-valued) Markov property: ``...the Markov-property criterion distinguishes descriptive levels at which the system exhibits a self-contained dynamics (`eigendynamics'), independent of details present at other levels.'' Emergent macroscopic processes are then identified with coarse-grainings which preserve the Markov property [``Markov partitions'' \citep{Addler:1998}]. We note that a Markovian coarse-grained macroscopic variable would automatically satisfy our criterion \eqref{eq:dians1} for dynamical independence.

Since low-level neural processes, and indeed neurophysiological recordings of these processes, do not naturally take the form of $1$st-order discrete-valued Markov processes, \citet{AllefeldAtmanspacherWackermann:2009} devise a discrete approximation scheme\footnote{Unfortunately, it is unclear how neural processes, which typically feature signal propagation delays, feedback over a range of time scales and medium- to long-range memory, might in general be well-represented by $1$st-order discrete-valued Markov processes at a fixed time increment.}. They then seek Markovian coarse-grainings of the discretised Markov model in the form of metastable macrostates \citep{OlivieriVares:2005}, and (putatively emergent) dynamics that transition between such macrostates at slow time scales compared to the underlying microscopic dynamics. This latter idea, more closely aligned with a thermodynamical perspective on coarse-graining \citep{Green:1952,JefferyEtal:2019}, seems worthy of further investigation in regard to dynamical independence.

\citet{ShaliziMoore:2003} consider the ``causal states'' of a system, defined as equivalence classes of state histories which yield the same conditional distribution over future states. The sequence of causal states defines a Markov process. A (coarse-grained) macroscopic process is then deemed emergent if its causal states self-predict ``more efficiently'' than the causal states of the microscopic process, where predictive efficiency of a process is measured in terms of the ratio of entropy rate to statistical complexity.

\citet{HoelEtal:2013} formulate a notion of causal emergence based on \emph{effective information} \citep{TononiSporns:2003}. Here, although macro is supervenient on micro, a coarse-grained macroscopic variable is deemed emergent to the extent that it leads to a gain in effective information. Effective information is calculated by comparing the distribution of prior states that could have caused a given current state (the ``causal distribution''), with the uniform distribution over the full repertoire of possible prior states. The KL-divergence of the causal distribution with respect to the uniform distribution is then averaged over the distribution of current states. The procedure is motivated by the Pearlian approach \citep{Pearl:2009} which identifies causation with the effects of counterfactual interventions (perturbations) on the system; the uniform (maximum entropy) distribution then stands as an injection of random perturbations. A drawback of effective information, however, is that it assumes the \emph{existence} of a uniform distribution of states, thus ruling out a large class of (in particular continuous-state) physical systems, for which the uniform distribution does not exist; and even if it exists, it is not clear that the EI will be transformation-invariant. It may also be argued that a uniform distribution over prior states is in any case a purely notional, unphysical construct, and that its deployment consequently fails to reflect causation ``as it actually happens'' -- that is, as stochastic dynamics play out over time. In a related approach, \citet{FristonEtal:2021} present a recursive partitioning of neuronal states based on effective connectivity graphs \citep{FristonEtal:2013} and Markov blankets \citep{Pearl:1998}, which they associate with emergent intrinsic brain networks at hierarchical spatiotemporal scales\footnote{Although described as a ``renormalisation group'' approach, it is never adequately explained why (or indeed whether) the dimensional reductions associated with partitioning should lead to self-similar dynamics at increasingly coarse scales.}.

\citet{Millidge:2021} presents a mathematical theory of abstraction which shares some commonalities with theories of emergence. An abstraction is considered as a set of ``summaries'' of a system which are sufficient to answer a specified set of ``queries'' regarding the time evolution of the system. Like macroscopic variables (in the broad sense), abstractions discard information about the system's detailed dynamics -- in this case such information as turns out to be irrelevant to the specific queries. It is proposed that the irrelevant information be considered via the \emph{maximum-entropy principle} \citep{Jaynes:1985}, whereby uncertainty about detailed system behaviour is maximised within the constraint of retention of the ability to answer the queries. Like dynamically-independent macrovariables, abstractions might be considered to have a ``life of their own'' insofar as they retain sufficient information to predict their own behaviours at a macroscopic level. In common with our approach, \citet{Millidge:2021} places an emphasis on \emph{data-driven discovery} of abstractions, by minimising their ``leakiness'' -- that is, their departure from accurate prediction of the associated macrophenomena (\cf~dynamical dependence). In contrast to dynamical independence, abstractions might be said to be driven by the agenda of the observer (in the form of specific queries), rather than, as in our case, unconstrained and intrinsic to the dynamical structure of the microsystem.

Finally, our approach is also clearly related to the general idea of dimensionality reduction in information theory, machine learning and beyond. Importantly, dynamical independence defines a very specific basis for dimensionality reduction, one which flows explicitly from the dynamics of the underlying microscopic system. This might be contrasted, for example, with principal components analysis (PCA), which is essentially determined by correlations within a dataset. In case the data derives from a dynamical process (\eg, econometric data, neuroimaging data, \etc), these correlations are \emph{contemporaneous}, and as such fail to reflect in full the temporal dynamics of the generative process.

\subsection{Relationship with autonomy} \label{sec:relaut}

A macroscopic process $Y$ that is dynamically-independent with respect to the microscopic process $X$ might well be described as ``autonomous of $X$''. We avoid this usage, though, because conventionally the term autonomy carries two distinct connotations \citep{BertschingerEtal:2008}: an autonomous process should not only be independent of external ``driving'' processes, but should also \emph{self-determine} its evolution over time \citep{Seth:2010}. As remarked in \secref{sec:emergence}, a dynamically-independent macroscopic variable need not fulfil the self-determination criterion; dynamical independence does not equate to autonomy (\cf~\secref{sec:relapp}, Granger autonomy/emergence). In the extreme case, a dynamically-independent macroscopic variable might in fact be completely random, as in the following trivial VAR($1$) example:
\begin{subequations}
\begin{align}
	X_{1,t} &= aX_{1,t-1} + bX_{2,t-1} + \eps_{1,t} \label{eq:minvar11} \\
	X_{2,t} &= \eps_{2,t} \label{eq:minvar12}
\end{align} \label{eq:minvar1}%
\end{subequations}
where $\eps_{1,t},\eps_{2,t}$ are uncorrelated white noises. Here the macroscopic (coarse-grained) white noise $Y_t = X_{2,t}$ is clearly dynamically-independent of the microscopic process $X_t$. Note, however, that a completely random macroscopic variable is not necessarily dynamically-independent: if we replace \eqref{eq:minvar12} with
\begin{equation}
	X_{2,t} = \eps_{2,t} + c\eps_{1,t-1} \label{eq:minvar12alt}
\end{equation}
then, while $Y_t = X_{2,t}$ is still a white noise, it is no longer dynamically-independent of $X_t$\footnote{$X_t$ is now VARMA($1,1$) rather than VAR($1$).}.

We consider this as a positive feature of our definition of dynamical independence: as per our \textit{ansatz} \eqref{txt:QQ}, if we discover that our microscopic system features a completely-random macroscopic variable at some scale, this tells us something useful about the system. We might even, via \eqref{eq:fsplit}, choose to ``factor out'' this embedded randomness in order to better reveal significant causal structure.

\subsection{Discovery of emergent macroscopic processes in neural systems} \label{sec:dineuro}

Notwithstanding that the generative mechanisms underlying neural processes may be highly nonlinear, linear modelling is routinely deployed for the functional analysis of neural systems via neurophysiological recordings (indeed, correlation statistics are associated with linear regression; see also our discussion in \secref{sec:cglss}). Granger causality based on VAR (and more recently state-space) modelling in particular is a popular technique for inference of directed functional connectivity \citep{Seth:gcneuro:2015,Barnett:ssgc:2015} from EEG, MEG and iEEG data\footnote{Granger-causal analysis of fMRI BOLD data, however, remains controversial, due to confounds related to slow sampling rates \citep{Seth:gcfmri:2013} and potentially also to the haemodynamic response function \citep[HRF;][]{Solo:2016}.}. The techniques described in \secref{sec:maxdi} may thus be applied directly to estimated state-space models for such data, to infer the emergence portrait of neural systems. Issues of scale (see \secref{sec:macglssn}) remain significant at this stage, but do not appear to be intractable.

While it may be tempting to draw analogies between dynamically-independent macrovariables in neural systems and functional network analyses, \eg, default-mode networks \citep{RaichleEtal:2001}, this would be misleading; a dynamically-independent macrovariable is not a static ``network'', but rather a macro-scale dynamical entity in its own right, emerging from interactions on the ``microscopic'' scale (in this case, the scale set by neural recording channels associated with ``small'' brain regions). A fascinating question for future empirical research, is whether specific emergent (dynamically-independent) macrovariables might be associated with (``neural correlates'' of) large-scale neural phenomena, such as behaviours, cognition, and specific states of, or disorders of, consciousness.

\subsection{Caveat} \label{sec:caveats}

Finally, a caveat: the underlying intuition behind any study of emergence for real-world systems, is that identifying emergent structure is likely to advance our understanding of the physical phenomena in question. While reasonable, this conclusion is not a given. Whether emergent dynamical structures turn out to be functionally relevant for explaining a particular system's behaviour will most often be an empirical question.

\subsection*{Acknowledgements}

We are grateful to the Dr. Mortimer and Theresa Sackler Foundation, which supports the Sackler Centre for Consciousness Science. The authors would also like to thank Nadine Spychala, Adam Barrett, Javier Galad\'i, Fernando Rosas, Pedro Mediano and Boki Milinkovic for useful discussions.

\clearpage
\appendix

\section{Proof of transitivity of dynamical independence} \label{sec:ditrans}

From property \eqref{eq:fsplit} we construct isomorphisms
\begin{subequations}
\begin{align}
	f \times u : \xstate \to \ystate \times \ustate \\
	g \times v : \ystate \to \zstate \times \vstate
\end{align}%
\end{subequations}
as in the diagram below
\begin{center}
\begin{tikzcd}
	\xstate \arrow[rd,"u"] \arrow[r,"f"] & \ystate \arrow[rd,"v"] \arrow[r,"g"] & \zstate\\
	& \ustate & \vstate
\end{tikzcd}
\end{center}
so that
\begin{equation}
	(g \circ f) \times (v \circ f) \times u : \xstate \to \zstate \times \vstate \times \ustate
\end{equation}
is an isomorphism. Setting $U_t = u(X_t)$, $V_t = v(Y_t) = v(f(X_t))$, under this isomorphism, we have $X_t \sim (Z_t,V_t,U_t)$ with $Y_t \sim (Z_t,V_t)$, and by \eqref{eq:displit}
\begin{subequations}
\begin{align}
	\teop(X \to Y \vbar E_t) &= \teop(U \to Z,V\vbar E_t) \label{eq:Txy} \\
	\teop(Y \to Z\vbar E_t) &= \teop(V \to Z\vbar E_t)    \label{eq:Tyz} \\
	\teop(X \to Z\vbar E_t) &= \teop(V,U \to Z\vbar E_t)  \label{eq:Txz}
\end{align} \label{eq:Txyz}%
\end{subequations}
The dynamical independence of $Y_t$ from $X_t$, and of $Z_t$ from $Y_t$ then become [\cf~\eqref{eq:tedd}]
\begin{subequations}
\begin{align}
	\enop(Z_t,V_t \vbar Z_t^-,V_t^-,E_t^-) & = \enop(Z_t,V_t \vbar Z_t^-,V_t^-,U_t^-,E_t^-) \label{eq:Hxy} \\
	\enop(Z_t \vbar Z_t^-,E_t^-) & = \enop(Z_t \vbar Z_t^-,V_t^-,E_t^-)                     \label{eq:Hyz}
\end{align} \label{eq:Hxyz}%
\end{subequations}
respectively, while
\begin{equation}
	\teop(X \to Z\vbar E_t) = \enop(Z_t \vbar Z_t^-,E_t^-) - \enop(Z_t \vbar Z_t^-,V_t^-,U_t^-,E_t^-) \label{eq:Hxz}
\end{equation}
Now by \eqref{eq:Hxy}, conditional on $(Z_t^-,V_t^-,E_t^-)$, the joint variable $(Z_t,V_t)$ is independent of $U_t^-$. Thus, again conditional on $(Z_t^-,V_t^-,E_t^-)$, the marginal $Z_t$ is itself independent of $U_t^-$. We thus have\footnote{Note that this conclusion holds for both discrete, and---with \emph{differential} entropy---for continuous-valued state.}
\begin{equation}
	\enop(Z_t \vbar Z_t^-,V_t^-,E_t^-) = \enop(Z_t \vbar Z_t^-,V_t^-,U_t^-,E_t^-) \label{eq:Hxym} \\
\end{equation}
which, together with \eqref{eq:Hyz} and \eqref{eq:Hxz} yields $\teop(X \to Z \vbar E_t) = 0$ and \eqref{eq:ditrans} holds as required.\hfill$\blacksquare$

\section{Granger causality} \label{sec:gc}

We begin by noting that the optimal linear prediction of $X_t$ in the least-squares sense is given by the conditional expectation $\cexpect{X_t}{X_t^-} = \sum_{k=1}^\infty A_k X_{t-k}$ \eqref{eq:AR}. The residual prediction errors are then just the innovations $\beps_t = X_t - \cexpect{X_t}{X_t^-}$, and in the formulation of \citet{Geweke:1982} the magnitude of the prediction error is quantified by the generalised variance \citep{Wilks:1932,Barrett:2010} $|\Sigma|$, where $\Sigma = \expect{\beps_t\beps_t^\trop}$ is the error covariance matrix.

Suppose now that the vector process $X_t$ is partitioned into two sub-processes $X_t = [X_{1t}^\trop\; X_{2t}^\trop]^\trop$. To specify the Granger causality $\gcop(X_2 \to X_1)$ we compare the prediction error $|\Sigma_{11}|$ of $X_{1t}$ predicted on the joint past $X_t^-$ of both itself and $X_{2t}$, with the prediction error $|\Sigma_{11}^\rrop|$ of  $X_{1t}$ predicted only on its own past $X_{1t}^-$; here the superscript ``$^\rrop$'' refers to the ``restricted'' VAR representation
\begin{equation}
	X_{1t} = \sum_{k=1}^\infty A^\rrop_k X_{1,t-k} + \beps_{1t}^\rrop \label{eq:rAR}
\end{equation}
and $\Sigma_{11}^\rrop = \expect{\beps_{1t}^\rrop\beps_{1t}^{\rrop\trop}}$ is the corresponding error covariance matrix. \citet{Geweke:1982} then defines the Granger causality as the log-ratio of generalised variances
\begin{equation}
	\gcop(X_2 \to X_1)= \log\frac{\big|\Sigma_{11}^\rrop\big|}{\big|\Sigma_{11} \big|} \label{eq:gc}
\end{equation}
which quantifies the degree to which the history of $X_{2t}$ enhances prediction of $X_{1t}$ beyond the degree to which $X_{1t}$ is predicted by its own history alone. We note that if the innovations are Gaussian, then the generalised variance $|\Sigma|$ is proportional to the likelihood function for $X_t$, so that \eqref{eq:gc} is a log-likelihood ratio, which under ergodic assumptions is asymptotically equivalent to the conditional entropy $\enop(X_t \vbar X_t^-)$ \citep{Barnett:teml:2012}; this circumscribes the relationship between Granger causality and transfer entropy [\cf~\eqref{eq:teddb}]. Under the classical ``large-sample theory'' \citep{NeymanPearson:1933,Wilks:1938,Wald:1943}, the log-likelihood ratio as a sample statistic furnishes asymptotic $F$- and $\chi^2$ tests for statistical inference on GC [but see also \citet{GutknechtBarnett:2019}].

Unlike transfer entropy, Granger causality may also be defined in the frequency domain: the spectral GC from $X_{2t}$ to $X_{1t}$ at angular frequency $\omega \in [0,2\pi]$ is given by \citep{Geweke:1982,Barnett:ssgc:2015}
\begin{equation}
	\sgcop(X_2 \to X_1;z) = \log\frac{|S_{11}(z)|}{\big|S_{11}(z) - H_{12}(z) \Sigma_{22|1} H_{12}^*(z) \big|} \label{eq:sgc}
\end{equation}
where $z = e^{-i\omega}$. Here $S(z)$ is the CPSD matrix \eqref{eq:cpsd}, $H(z)$ the transfer function \eqref{eq:MA}, and $\Sigma_{22|1} = \Sigma_{22} - \Sigma_{21} \Sigma_{11}^{-1} \Sigma_{12}$ a partial covariance matrix. Spectral GC averages across the broadband frequency range $[0,2\pi]$ to yield time-domain GC \citep{Geweke:1982}:
\begin{equation}
	\gcop(X_2 \to X_1) = \frac1{2\pi} \int_0^{2\pi} \sgcop\big(X \to Y;e^{-i\omega}\big)\,d\omega \label{eq:sgcint}
\end{equation}
Given a frequency band $[\omega_1,\omega_2] \subseteq [0,2\pi]$, we may define the ``band-limited'' (time-domain) Granger causality \citep{Barnett:gcfilt:2011}
\begin{equation}
	\gcop(X_2 \to X_1;\omega_1,\omega_2) = \frac1{\omega_2-\omega_1} \int_{\omega_1}^{\omega_2} \sgcop\big(X_2 \to X_1;e^{-i\omega}\big)\,d\omega \label{eq:blgc}
\end{equation}
which may be interpreted as the information transfer from $X_2$ to $X_1$ associated with frequencies $\omega_1 \le \omega \le \omega_2$.

\subsection{Granger causality for linear state-space systems} \label{sec:ssgc}

Suppose that the observation process $X_t$ for an innovations-form SS model \eqref{eq:iss} is partitioned as above into two sub-processes. Following \citet{Barnett:ssgc:2015}, we show how $\gcop(X_2 \to X_1)$ may be calculated from the ISS parameters. The utility of innovations form \eqref{eq:iss} for GC analysis, is that the innovations $\beps_t$ are precisely the residual error terms of the predictive VAR representation \eqref{eq:AR} for $X_t$, which feature in the expression for Granger causality.

Given \eqref{eq:iss}, $X_{1t}$ satisfies the SS model, now no longer in innovations form:
\begin{subequations}
\begin{align}
	Z_{t+1}  &= A\;Z_t \,+ K \beps_t \label{eq:rissz} \\
	X_{1t} &= C_1 Z_t + \phantom{K}\beps_{1t} \label{eq:rissx}
\end{align} \label{eq:riss}%
\end{subequations}
where $C$, $\beps_t$ and $\Sigma$ are partitioned concordantly with $X_{1t},X_{2t}$. The joint noise covariance matrices for the SS \eqref{eq:riss} are given by
\begin{equation}
	Q = K \Sigma K^\trop\,, \qquad
	R = \Sigma_{11}\,, \qquad
	S = K \Sigma_{*1} \label{eq:rrescovs}
\end{equation}
where $\Sigma_{*1} = [\Sigma_{11}\;\Sigma_{12}]^\trop$ and, converting to innovations form, from \eqref{eq:p2iSig} we have
\begin{equation}
	\Sigma_{11}^\rrop = C_1 P C_1^\trop + \Sigma_{11}
\end{equation}
where $P$ is the unique stabilising solution\footnote{Since $X_t$ is purely nondeterministic, so is $X_{1t}$, so that $R = \Sigma_{11}$ is positive-definite; $X_{1t}$ is also miniphase, so that all conditions for a unique stabilising solution are met; see \citet{Barnett:ssgc:2015,Solo:2016}.} of the ``reduced'' DARE [\cf~\eqref{eq:dare}]
\begin{equation}
	P =  APA^\trop + Q - \big(APC_1^\trop + S\big) \big(C_1PC_1^\trop + R\big)^{-1} \big(C_1PA^\trop + S^\trop\big) \label{eq:rdare}
\end{equation}
with $Q,R,S$ as in \eqref{eq:rrescovs}, and the GC \eqref{eq:gc} may thus be calculated. The condition for vanishing $\gcop(X_2 \to X_1)$ is \citet[eq.~17 \& \textit{ff.}]{Barnett:ssgc:2015}
\begin{equation}
	\gcop(X_2 \to X_1) = 0 \iff C_1 A^k K_2 \equiv 0 \quad\text{for}\quad k = 0,1,\ldots,r-1 \label{eq:ssgc0}
\end{equation}

In the spectral domain, the formula \eqref{eq:sgc} applies, with the transfer function as in \eqref{eq:MA}, and the CPSD specified by \eqref{eq:cpsd}. Note that in the unconditional case presented here, there is no need to solve the DARE \eqref{eq:rdare}; the conditional spectral GC (required in particular if there is an environmental process $\envi_t$) is somewhat more complex \citep{Barnett:ssgc:2015}, and requires solution of a DARE.

\subsection{Granger causality for finite-order VAR systems} \label{sec:gcar}

Although a special case of state-space (equivalently VARMA) models, here we show that for finite-order pure-autoregressive models we may achieve a reduction in computational complexity.

We suppose that a VAR($p$) model, $p < \infty$, is given by
\begin{equation}
	X_t = \sum_{k=1}^p A_k X_{t-k} + \beps_t\,, \qquad \Sigma = \expect{\beps_t\beps_t^\trop} \label{eq:ar}
\end{equation}
with $n \times n$ coefficients matrices $A_1,\ldots,A_p$ and residuals covariance matrix $\Sigma$. We may specify an equivalent (innovations-form) state-space model by
\begin{subequations}
\begin{align}
	Z_{t+1} &= A Z_t + K\beps_t \label{eq:arissz} \\
	X_t     &= C Z_t + \phantom{K}\beps_t \label{eq:arissy}
\end{align} \label{eq:ariss}%
\end{subequations}
where $A$ is the $pn \times pn$ ``companion matrix'' \citep{HandD:2012}
\begin{equation}
	A =
	\begin{bmatrix}
		A_1 & A_2 & \ldots & A_{p-1} & A_p \\
		I   & 0   & \ldots & 0       & 0   \\
		0   & I   & \ldots & 0       & 0   \\
		\vdots & \vdots & \ddots & \vdots & \vdots   \\
		0   & 0   & \ldots & I       & 0
	\end{bmatrix}
\end{equation}
$C$ the $n \times pn$ matrix
\begin{equation}
	C =
	\begin{bmatrix}
		A_1 & A_2 & \ldots & A_{p-1} & A_p
	\end{bmatrix}
	\label{eq:AC}
\end{equation}
and $K$ the $pn \times n$ matrix
\begin{equation}
	K =
	\begin{bmatrix}
		I & 0 & \ldots & 0
	\end{bmatrix}^\trop
\end{equation}
The VAR($p$) is stable iff the companion matrix $A$ is stable, and is always miniphase.

We suppose again that $X_t$ is partitioned into sub-processes $X_{1t},X_{2t}$. We may then calculate the Granger causality $\gcop(X_2 \to X_1)$ as already described in \apxref{sec:ssgc}. Note that the DARE \eqref{eq:rdare} is a $pn \times pn$ matrix equation. In \citet[Appendix~B]{GutknechtBarnett:2019} it is shown that for VAR systems, the DARE may be reduced to $pn_2 \times pn_2$ dimensions, where $n_2$ is the dimension of $X_{2t}$. Specifically, the reduced model residuals covariance matrix $\Sigma_{11}^\rrop$ in the expression \eqref{eq:gc} for the Granger causality $\gcop(X_2 \to X_1)$ is given by
\begin{equation}
		\Sigma_{11}^\rrop = C_{12} P C_{12}^\trop + \Sigma_{11} \label{eq:rred}
\end{equation}
where $P$ is the solution of the $pn_2 \times pn_2$ DARE
\begin{equation}
		P = A_{22} P A_{22}^\trop + Q_{22} - (A_{22} P C_{12}^\trop + S_{21}) (C_{12} P C_{12}^\trop + R)^{-1} (A_{22} P C_{12}^\trop + S_{21}^\trop)^\trop \label{eq:ardare1}
\end{equation}
with
\begin{equation}
	A_{22} =
	\begin{bmatrix}
		A_{1,22} & A_{2,22} & \ldots & A_{p-1,22} & A_{p,22} \\
		I   & 0   & \ldots & 0       & 0   \\
		0   & I   & \ldots & 0       & 0   \\
		\vdots & \vdots & \ddots & \vdots & \vdots   \\
		0   & 0   & \ldots & I       & 0
	\end{bmatrix}\,, \qquad\quad
	C_{12} =
	\begin{bmatrix}
		A_{1,12} & A_{2,12} & \ldots & A_{p-1,12} & A_{p,12}
	\end{bmatrix}
	\label{eq:A22C12}
\end{equation}
and
\begin{equation}
	Q_{22} =
	\begin{bmatrix}
		\Sigma_{22} & 0 & \ldots & 0 \\
		0      & 0 & \ldots & 0 \\
		\vdots & \vdots & \ddots & \vdots \\
		0      & 0 & \ldots & 0
	\end{bmatrix}\,, \qquad\quad
	R = \Sigma_{11}\,, \qquad\quad
	S_{21} =
	\begin{bmatrix}
		\Sigma_{21} \\ 0 \\ \vdots \\ 0
	\end{bmatrix}
	\label{eq:Q22RS21}
\end{equation}
 The condition for vanishing $\gcop(X_2 \to X_1)$ is \citet[eq.~17 \& \textit{ff.}]{Barnett:ssgc:2015}
\begin{equation}
	\gcop(X_2 \to X_1) = 0 \iff A_{k,12} \equiv 0 \quad\text{for}\quad k = 1,\ldots,p \label{eq:argc0}
\end{equation}
For spectral GC, the formula \eqref{eq:sgc} applies, with transfer function
\begin{equation}
	H(z) = \bracr{I - \sum_{k=1}^p A_k z^k}^{-1} \label{eq:artf}
\end{equation}

\section{Analytic optimisation of state-space dynamical dependence} \label{sec:macglssa}

To minimise $\gcop(X \to Y) = \log|LVL^\trop|$ (eq.~\ref{eq:ssdd}) over the set of $m \times n$ matrices\footnote{In what follows, latin subscripts index $\reals^n$ coordinates, greek subscripts $\reals^m$ coordinates.} $L = (L_{\alpha i})$ under the orthonormality constraint $LL^\trop = I$, we introduce Lagrange multipliers $\Lambda = (\lambda_{\alpha\beta})$, ($\Lambda$ is $m \times m$ symmetric) and solve simultaneously
\begin{subequations}
\begin{align}
	\nabla  \log|LVL^\trop| &= \nabla  \trace{\Lambda LL^\trop} \label{eq:ddlgra} \\
	LL^\trop &= I \label{eq:ddlgrb}
\end{align} \label{eq:ddlgr}%
\end{subequations}
where $V = V(L) = CP(L)C^\trop+I$ with $P(L)$ the solution of the DARE \eqref{eq:dddares}, and for a tensor $\psi_{jk\ell\ldots}$ $\nabla\!\psi$ denotes the tensor with entries $\psi_{jk\ell\ldots ,\alpha i} = \displaystyle \frac{\partial\psi_{jk\ell\ldots}}{\partial L_{\alpha i}}$. We may calulate that
\begin{subequations}
\begin{align}
	\nabla \log|LVL^\trop| &= 2LQV + 2R \label{eq:ddlaga} \\
	\nabla \trace{\Lambda LL^\trop} &= 2\Lambda L \label{eq:ddlagb}
\end{align} \label{eq:ddlag}%
\end{subequations}
where we have set
\begin{subequations}
\begin{align}
	Q & = L^\trop\big(LVL^\trop\big)^{-1} L \\
	R &= \tfrac12 \trace{Q \cdot \nabla V}
\end{align} %
\end{subequations}
Here $Q$ is $n \times n$, and $\trace{Q \cdot \nabla V}$ is to be understood as the $m \times n$ matrix with entries $\sum_{j,k} Q_{jk} V_{kj,\alpha i}$. Equations \eqref{eq:ddlgr} and \eqref{eq:ddlag} then yield
\begin{equation}
	LQV + R	= \Lambda L
\end{equation}
Multiplying both sides on the right by $L^\trop$, and noting that  $LQVL^\trop = I$ yields the Lagrange multiplier
\begin{equation}
	\Lambda = I + RL^\trop
\end{equation}
and the optimisation equations may thus be written
\begin{equation}
	R(I-L^\trop L) = L(I-QV) \label{eq:ddopt1}
\end{equation}
to be solved under the constraint $LL^\trop = I$, where all derivatives of the DARE solution $P(L)$ with respect to $L$ are now collected in the LHS term $R(L)$. Note that if $LL^\trop = I$ then $I-L^\trop L$ will be singular, with rank $n-m$. To solve \eqref{eq:ddopt1} requires $\nabla V = C \cdot \nabla P \cdot C^\trop$, where $\nabla P$ may (in principle) be derived by partial differentiation of the DARE \eqref{eq:dddares} with respect to the $L_{\alpha i}$; \eqref{eq:ddopt1} then becomes a set of highly-nonlinear partial differential equations, where the terms $P$ and $\nabla P$ are only defined implicitly.  The calculation thus appears, at this stage, intractable.

\section{Dynamical independence for flows} \label{sec:hamdyn}

A smooth (local) flow $\xi(\bx,t)$ satisfying  \eqref{eq:flow} may be considered as a \emph{vector field} on $\reals^n$, with associated \emph{gradient operator}
\begin{equation}
	\nabla_{\!\xi} = \sum_{i=1}^n g_i \frac\partial{\partial x_i}
\end{equation}
where the associated ODE is $\dot\bx(t) = g(\bx)$, and for any function $\phi(\bx)$ on $\reals^n$ we have
\begin{equation}
	\frac\partial{\partial t} \phi\big(\xi(\bx,t)\big) = \nabla_{\!\xi} \phi\big(\xi(\bx,t)\big) \label{eq:Fdel}
\end{equation}
for all $\bx,t$. As described in \secref{sec:flow}, a macroscopic variable $\by(t) = f\big(\bx(t)\big)$, where $f : \reals^n \to \reals^m$ is a coarse-graining\footnote{The mapping $f(\bx)$ is (locally) surjective wherever the Jacobean matrix $\nabla\!f(\bx) = \left[\frac{\partial f_\alpha}{\partial x_i}\right]$ has full rank $ = m$.}, is deemed dynamically-independent of the process $\bx(t)$ if it is itself described by a flow; that is, if there is a flow $\eta : \reals^m \times \reals \to \reals^m$ such that \eqref{eq:ctdi} is satisfied. Differentiating \eqref{eq:ctdi} with respect to $t$ and setting $t = 0$, we find that $\by(t) = f\big(\bx(t)\big)$ is dynamically-independent with respect to $\bx(t)$ iff
\begin{equation}
	\nabla_{\!\xi}f(\bx) = h\big(f(\bx)) \label{eq:ctdicond}
\end{equation}
for all $\bx$, for some mapping $h : \reals^m \to \reals^m$; the associated ODE for $\by(t)$ is then $\dot\by(t) = h(\by)$.

We wish to find a general form for dynamically-independent coarse-grainings. We thus seek functions $f : \reals^n \to \reals^m$ which satisfy \eqref{eq:ctdicond} for some $h : \reals^m \to \reals^m$. Excluding the trivial flow $\xi(\bx,t) = \bx$ [where $g(\bx) \equiv 0$], as a preliminary step we seek solutions to the PDEs
\begin{subequations}
\begin{align}
	\nabla_{\!\xi}\tau(\bx) &= 1 \label{eq:fpar} \\
	\nabla_{\!\xi}u(\bx) &= 0 \label{eq:finv}
\end{align} \label{eq:tildegh1}%
\end{subequations}
for scalar functions $\tau(\bx),u(\bx)$. From  from \eqref{eq:Fdel}, we see that solutions of \eqref{eq:finv} are \emph{invariants} of the flow $\xi$---that is, $u\big(\xi(\bx,t)\big)$ does not vary with $t$---while solutions of \eqref{eq:fpar} ``parametrise time'' along the trajectories of $\xi$, in the sense that $\tau\big(\xi(\bx,t)\big)$ changes at a constant unit rate with $t$. If the flow $\xi(\bx,t)$ is known explicitly, invariants may be obtained analytically by eliminating $t$ between pairs $\xi_i(\bx,t), \xi_j(\bx,t)$, $i \ne j$. It will thus in general (at least locally on $\reals^n$) be possible to find a basis set $\bu(\bx) = \big(u_1(\bx),\ldots,u_{n-1}(\bx)\big)$ of $n-1$ \emph{functionally-independent} invariants\footnote{That is, the Jacobean matrix $\nabla\!u = \left[\frac{\partial u_j}{\partial x_i}\right]$ has full rank $ = n-1$.}, such that any invariant of the flow is of the form $U\big(\bu(\bx)\big)$.

Given a set of $n-1$ invariants $\bu = (u_1,\ldots,u_{n-1})$, to solve \eqref{eq:fpar} let $v(\bx)$ be any scalar function such that $\bx \mapsto \big(v(\bx),\bu(\bx)\big)$ defines a nonsingular transformation of $\reals^n$. In the new coordinate system $(v,u_1,\ldots,u_{n-1})$ we may calculate
\begin{equation}
	\nabla_{\!\xi} = \gamma(v,\bu) \frac\partial{\partial v}
\end{equation}
where $\gamma(\bu,v)$ is defined implicitly by $\gamma(v(\bx),\bu(\bx)) = \nabla_{\!\xi}v(\bx)$. \eqref{eq:fpar} then becomes
\begin{equation}
	\gamma(v,\bu) \frac{\partial\tau}{\partial v} = 1 \label{eq:gammavu}
\end{equation}
Setting
\begin{equation}
	\theta(v,\bu) = \int \frac{dv}{\gamma(v,\bu)} \qquad\text{(indefinite integral)} \label{eq:VINT}
\end{equation}
from \eqref{eq:gammavu}, the general solution to \eqref{eq:fpar} is thus
\begin{equation}
	\tau(\bx) = \theta\big(v(\bx),\bu(\bx)\big) + U(\bx) \label{eq:taux}
\end{equation}
where $U(\bx)$ is an arbitrary invariant. The choice of particular $v(\bx)$ is not significant, insofar as we may verify\footnote{Via a change of variable $v' = \upsilon(v,\bu)$.} that $\theta\big(v(\bx),\bu(\bx)\big)$ is unique up to an additive function of $\bu(\bx)$ alone, which may be absorbed in $U(\bx)$. In practice $v(\bx)$ may be chosen for convenience of evaluation of the indefinite integral in \eqref{eq:VINT}. In the coordinate system $(\tau,\bu)$, we have $\nabla_{\!\xi} = \frac\partial{\partial\tau}$; intuitively, the transformation $\bx \mapsto \big(\tau(\bx),\bu(\bx)\big)$ ``flattens out'' the flow, so that points in $\reals^n$ are transported at unit rate along straight-line trajectories parallel to the $\tau$-axis.

Returning to the general solution to \eqref{eq:ctdicond}---\ie, finding all dynamically-independent coarse-graining maps $f : \reals^n \to \reals^m$ for the flow $\xi$---we take first the case where the coarse-grained flow $\eta$ is non-trivial (we continue to assume that $\xi$ is non-trivial). Then, from the above analysis, we can always find transformations of $\reals^n$ and $\reals^m$ that (at least locally) ``flatten out'' the respective flows $\xi,\eta$ as described above. Under these transformations, the condition \eqref{eq:ctdicond} for dynamical-independence becomes
\begin{subequations}
\begin{align}
	\frac{\partial f_1}{\partial \tau} &= 1 \\[0.5em]
	\frac{\partial f_\alpha}{\partial \tau} &= 0\,, \qquad \alpha = 2,\ldots,m
\end{align}%
\end{subequations}
so that, transforming back to the original coordinates, the general form of a dynamically-independent coarse-graining is
\begin{equation}
	f(\bx) = \Psi\big(\tau(\bx),\bu(\bx)\big) \label{eq:fsol}
\end{equation}
with $\bu(\bx) = \big(u_1(\bx),\ldots,u_{m-1}(\bx)\big)$ a set of $m-1$ functionally-independent invariants, $\tau(\bx)$ as in \eqref{eq:taux}, and $\Psi(\by)$ an arbitrary diffeomorphism of $\reals^m$. In the case where the coarse-grained flow $\eta(\by,t)$ is trivial, it is easy to see that $f(\bx)$ must itself comprise a set of $m$ functionally-independent invariants.

\subsection{Worked example} \label{sec:hamdynwe}

Consider the system of ODEs on $\reals^3$ defined by
\begin{subequations}
\begin{align}
	\dot x_1 &= -x_2 \\
	\dot x_2 &= \phantom-x_1 \\
	\dot x_3 &= 2x_1 x_2
\end{align} \label{eq:wx1odes}%
\end{subequations}
We have $g(\bx) = (-x_2,x_1,2x_1x_2)$, so that
\begin{equation}
	\nabla_{\!\xi} = -x_2 \frac\partial{\partial{x_1}} + x_1 \frac\partial{\partial{x_2}} + 2x_1 x_2 \frac\partial{\partial{x_3}} \label{eq:xigrad}
\end{equation}
and the flow corresponding to \eqref{eq:wx1odes} is given by\footnote{Change to polar coordinates in $x_1,x_2$, then integrate to solve for $\xi_3$.}
\begin{subequations}
\begin{align}
	\xi_1(\bx,t) &= x_1\cos t - x_2\sin t \label{eqa:wx1flow} \\
	\xi_2(\bx,t) &= x_1\sin t + x_2\cos t \label{eqb:wx1flow} \\
	\xi_3(\bx,t) &= x_3 + 2x_1x_2\cos t\sin t + \big(x_1^2-x_2^2\big) \sin^2 t \label{eqc:wx1flow}
\end{align} \label{eq:wx1flow}%
\end{subequations}
\begin{figure}
	\begin{center}
	\includegraphics[scale=0.6]{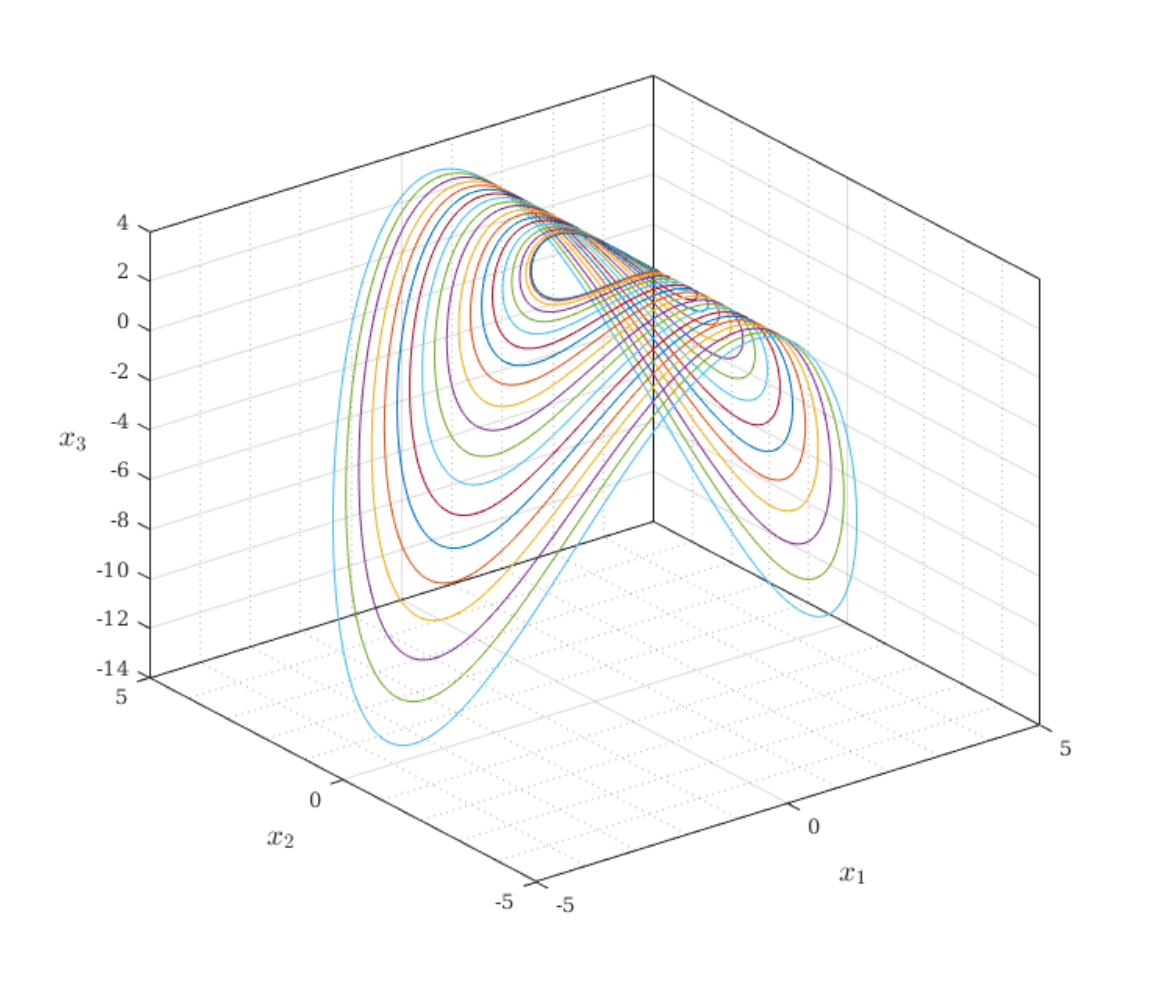}
	\end{center}
	\caption{Trajectories of the flow \eqref{eq:wx1flow} for $x_1 = 1$, $x_3 = 3$, and $x_2$ varying in the range $0\text{--}4$.} \label{fig:pringles}
\end{figure}
See \figref{fig:pringles}. An invariant $u(\bx)$ must satisfy \eqref{eq:finv}. We may confirm\footnote{This is essentially the ``method of characteristics''.} that
\begin{subequations}
\begin{align}
	u_1 &= x_1^2 + x_3 \label{eqa:u1inv} \qquad(\text{eliminate } t \text{ between eqs. \ref{eqa:wx1flow} and \ref{eqc:wx1flow}}) \\
	u_2 &= x_2^2 - x_3 \label{eqb:u2inv} \qquad(\text{eliminate } t \text{ between eqs. \ref{eqb:wx1flow} and \ref{eqc:wx1flow}})
\end{align} \label{eq:wx1invs}%
\end{subequations}
are functionally-independent invariants and for simplicity we choose $v = x_3$, which is functionally-independent of $u_1,u_2$. We have
\begin{equation}
	\gamma = \nabla_{\!\xi} v = 2x_1x_2 = 2\sqrt{(u_1-v)(u_2+v)}
\end{equation}
so that
\begin{equation}
	\theta = \int \frac{dv}{2\sqrt{(u_1-v)(u_2+v)}}
	= -\frac12 \sin^{-1}\left(\frac{u_1-u_2-2v}{u_1+u_2}\right)
	= -\frac12 \sin^{-1}\left(\frac{x_1^2-x_2^2}{x_1^2+x_2^2}\right)
\end{equation}
and the general solution to \eqref{eq:fpar} is then
\begin{equation}
	\tau(\bx) =  - \frac12 \sin^{-1}\left(\frac{x_1^2-x_2^2}{x_1^2+x_2^2}\right) + U\big(x_1^2+x_3,x_2^2-x_3\big)
\end{equation}
where $U(u_1,u_2)$ is an arbitrary function of two variables. The dynamically-independent coarse-grainings are then obtained from \eqref{eq:fsol}.

\clearpage

\bibliographystyle{apalike}
\bibliography{DynamicalIndependence}

\begin{thebibliography}{}

\bibitem[Adler, 1998]{Addler:1998}
Adler, R.~L. (1998).
\newblock Symbolic dynamics and {M}arkov partitions.
\newblock {\em Bull. Amer. Math. Soc.}, 35(1):1--56.

\bibitem[Allefeld et~al., 2009]{AllefeldAtmanspacherWackermann:2009}
Allefeld, C., Atmanspacher, H., and Wackermann, J. (2009).
\newblock Mental states as macrostates emerging from brain electrical dynamics.
\newblock {\em Chaos}, 19(1):015102.

\bibitem[Arnold, 1978]{Arnold:1978}
Arnold, V.~I. (1978).
\newblock {\em Mathematical Methods of Classical Mechanics}.
\newblock Springer-Verlag, New York, NY.

\bibitem[Barnett et~al., 2009]{Barnett:tegc:2009}
Barnett, L., Barrett, A.~B., and Seth, A.~K. (2009).
\newblock Granger causality and transfer entropy are equivalent for {G}aussian
  variables.
\newblock {\em Phys. Rev. Lett.}, 103(23):0238701.

\bibitem[Barnett and Bossomaier, 2013]{Barnett:teml:2012}
Barnett, L. and Bossomaier, T. (2013).
\newblock Transfer entropy as a log-likelihood ratio.
\newblock {\em Phys. Rev. Lett.}, 109(13):0138105.

\bibitem[Barnett and Seth, 2011]{Barnett:gcfilt:2011}
Barnett, L. and Seth, A.~K. (2011).
\newblock Behaviour of {G}ranger causality under filtering: {T}heoretical
  invariance and practical application.
\newblock {\em J. Neurosci. Methods}, 201(2):404--419.

\bibitem[Barnett and Seth, 2014]{Barnett:mvgc:2014}
Barnett, L. and Seth, A.~K. (2014).
\newblock The {MVGC} multivariate {G}ranger causality toolbox: {A} new approach
  to {G}ranger-causal inference.
\newblock {\em J. Neurosci. Methods}, 223:50--68.

\bibitem[Barnett and Seth, 2015]{Barnett:ssgc:2015}
Barnett, L. and Seth, A.~K. (2015).
\newblock Granger causality for state-space models.
\newblock {\em Phys. Rev. E (Rapid Communications)}, 91(4):040101(R).

\bibitem[Barnett and Seth, 2017]{Barnett:downsample:2017}
Barnett, L. and Seth, A.~K. (2017).
\newblock Detectability of {G}ranger causality for subsampled continuous-time
  neurophysiological processes.
\newblock {\em J. Neurosci. Methods}, 275:93--121.

\bibitem[Barrett, 2015]{Barrett:2015}
Barrett, A.~B. (2015).
\newblock Exploration of synergistic and redundant information sharing in
  static and dynamical {G}aussian systems.
\newblock {\em Phys. Rev. E}, 91:052802.

\bibitem[Barrett et~al., 2010]{Barrett:2010}
Barrett, A.~B., Barnett, L., and Seth, A.~K. (2010).
\newblock Multivariate {G}ranger causality and generalized variance.
\newblock {\em Phys. Rev. E}, 81(4):041907.

\bibitem[Bays, 2009]{Bays:2009}
Bays, C. (2009).
\newblock Gliders in cellular automata.
\newblock In Meyers, R.~A., editor, {\em Encyclopedia of Complexity and Systems
  Science}, pages 4240--4249. Springer, New York, NY.

\bibitem[Bedau, 1997]{Bedau:1997}
Bedau, M.~A. (1997).
\newblock Weak emergence.
\newblock {\em No\^us}, 31(11):375--399.

\bibitem[Bell, 2015]{SEP-AOC:2015}
Bell, J.~L. (2015).
\newblock The axiom of choice.
\newblock In Zalta, E.~N., editor, {\em The Stanford Encyclopedia of
  Philosophy}. The Metaphysics Research Lab, Center for the Study of Language
  and Information, Stanford University, Stanford, CA. Online:
  \url{https://plato.stanford.edu/entries/axiom-choice/}.

\bibitem[Bertschinger et~al., 2006]{BertschingerEtal:2006}
Bertschinger, N., Olbrich, E., Ay, N., and Jost, J. (2006).
\newblock Information and closure in systems theory.
\newblock In {\em Explorations in the Complexity of Possible Life. Proceedings
  of the 7th German Workshop of Artificial Life}, pages 26--28.

\bibitem[Bertschinger et~al., 2008]{BertschingerEtal:2008}
Bertschinger, N., Olbrich, E., Ay, N., and Jost, J. (2008).
\newblock Autonomy: {A}n information theoretic perspective.
\newblock {\em Biosystems}, 91(2):331--345.

\bibitem[Bossomaier et~al., 2016]{BossomaierEtal:2016}
Bossomaier, T., Barnett, L., Harr\'e, M., and Lizier, J.~T. (2016).
\newblock {\em An Introduction to Transfer Entropy: Information Flow in Complex
  Systems}.
\newblock Springer International Publishing, Switzerland.

\bibitem[Botev et~al., 2013]{BotevEtal:2013}
Botev, Z.~I., Kroese, D.~P., Rubinstein, R.~Y., and L’{E}cuyer, P. (2013).
\newblock The cross-entropy method for optimization.
\newblock In Rao, C.~R. and Govindaraju, V., editors, {\em Handbook of
  Statistics}, volume~31 of {\em Handbook of Statistics}, chapter~3, pages
  35--59. Elsevier.

\bibitem[Cavagna et~al., 2013]{CavagnaEtal:2013}
Cavagna, A., {Duarte Queir\'os}, S.~M., Giardina, I., Stefanini, F., and Viale,
  M. (2013).
\newblock Diffusion of individual birds in starling flocks.
\newblock {\em Proceedings of the Royal Society B: Biological Sciences},
  280(1756):20122484.

\bibitem[Chang et~al., 2020]{ChangEtal:2020}
Chang, A. Y.~C., Biehl, M., Yu, Y., and Kanai, R. (2020).
\newblock Information closure theory of consciousness.
\newblock {\em Front. Psychol.}, 11(1504).

\bibitem[Chicharro et~al., 2018]{ChicharroEtal:2018}
Chicharro, D., Pica, G., and Panzeri, S. (2018).
\newblock The identity of information: how deterministic dependencies constrain
  information synergy and redundancy.
\newblock {\em Entropy}, 20(3).

\bibitem[Cohen, 1911]{Cohen:1911}
Cohen, A. (1911).
\newblock {\em An introduction to the Lie Theory of One-Parameter Groups, with
  Applications to the Solution of Differential Equations}.
\newblock D. C. Heath \& Co., New York, NY.

\bibitem[Cover and Thomas, 1991]{CoverThomas:1991}
Cover, T.~M. and Thomas, J.~A. (1991).
\newblock {\em Elements of information theory}.
\newblock Wiley-Interscience, New York.

\bibitem[Davidson, 1980]{Davidson:1980}
Davidson, D. (1980).
\newblock Mental events.
\newblock In {\em Essays on Actions and Events}, chapter~11. Oxford University
  Press, New York, NY.

\bibitem[Dhamala et~al., 2008]{Dhamala:2008a}
Dhamala, M., Rangarajan, G., and Ding, M. (2008).
\newblock Estimating {G}ranger causality from {F}ourier and wavelet transforms
  of time series data.
\newblock {\em Phys. Rev. Lett.}, 100:018701.

\bibitem[Doob, 1953]{Doob:1953}
Doob, J. (1953).
\newblock {\em Stochastic Processes}.
\newblock John Wiley, New York.

\bibitem[Edelman et~al., 1998]{EdelmanEtal:1998}
Edelman, A., Arias, T., and Smith, S.~T. (1998).
\newblock The geometry of algorithms with orthogonality constraints.
\newblock {\em SIAM J. Matrix Anal. Appl.}, 20(2):303--353.

\bibitem[Friston et~al., 2013]{FristonEtal:2013}
Friston, K., Moran, R., and Seth, A.~K. (2013).
\newblock Analysing connectivity with {G}ranger causality and dynamic causal
  modelling.
\newblock {\em Curr. Opin. in Neurobiol.}, 23(2):172--178.
\newblock Macrocircuits.

\bibitem[Friston et~al., 2021]{FristonEtal:2021}
Friston, K.~J., Fagerholm, E.~D., Zarghami, T.~S., Parr, T., Hip\'olito, I.,
  Magrou, L., and Razi, A. (2021).
\newblock Parcels and particles: {M}arkov blankets in the brain.
\newblock {\em Network Neuroscience}, 5(1):211--251.

\bibitem[Geweke, 1982]{Geweke:1982}
Geweke, J. (1982).
\newblock Measurement of linear dependence and feedback between multiple time
  series.
\newblock {\em J. Am. Stat. Assoc.}, 77(378):304--313.

\bibitem[Geweke, 1984]{Geweke:1984}
Geweke, J. (1984).
\newblock Measures of conditional linear dependence and feedback between time
  series.
\newblock {\em J. Am. Stat. Assoc.}, 79(388):907--915.

\bibitem[Granger, 1963]{Granger:1963}
Granger, C. W.~J. (1963).
\newblock Economic processes involving feedback.
\newblock {\em Inform. Control}, 6(1):28--48.

\bibitem[Granger, 1969]{Granger:1969}
Granger, C. W.~J. (1969).
\newblock Investigating causal relations by econometric models and
  cross-spectral methods.
\newblock {\em Econometrica}, 37:424--438.

\bibitem[Green, 1952]{Green:1952}
Green, M.~S. (1952).
\newblock Markoff random processes and the statistical mechanics of
  time‐dependent phenomena.
\newblock {\em J. Chem. Phys.}, 20(8):1281--1295.

\bibitem[Gutknecht and Barnett, 2019]{GutknechtBarnett:2019}
Gutknecht, A.~J. and Barnett, L. (2019).
\newblock Sampling distribution for single-regression {G}ranger causality
  estimators.
\newblock {\em arXiv}, 1911.09625 [math.ST].

\bibitem[Hamilton, 1994]{Hamilton:1994}
Hamilton, J.~D. (1994).
\newblock {\em Time Series Analysis}.
\newblock Princeton University Press, Princeton, NJ.

\bibitem[Hannan and Deistler, 2012]{HandD:2012}
Hannan, E.~J. and Deistler, M. (2012).
\newblock {\em The Statistical Theory of Linear Systems}.
\newblock SIAM, Philadelphia, PA, USA.

\bibitem[Hansen et~al., 2015]{HansenEtal:2015}
Hansen, N., Arnold, D.~V., and Auger, A. (2015).
\newblock Evolution strategies.
\newblock In Kacprzyk, J. and Pedrycz, W., editors, {\em Springer Handbook of
  Computational Intelligence}, pages 871--898. Springer, Berlin, Heidelberg.

\bibitem[Helgason, 1978]{Helgason:1978}
Helgason, S. (1978).
\newblock {\em Differential Geometry, Lie Groups, and Symmetric Spaces}.
\newblock Academic Press, New York, NY.

\bibitem[Hlav\'a\v{c}kov\'a-Schindler, 2011]{HlavackovaSchindler:2011}
Hlav\'a\v{c}kov\'a-Schindler, K. (2011).
\newblock Equivalence of {G}ranger causality and transfer entropy: {A}
  generalization.
\newblock {\em Appl. Math. Sci.}, 5(73):3637--3648.

\bibitem[Hoel et~al., 2013]{HoelEtal:2013}
Hoel, E.~P., Albantakis, L., and Tononi, G. (2013).
\newblock Quantifying causal emergence shows that macro can beat micro.
\newblock {\em P. Natl. Acad. Sci. USA}, 110(49):19790--19795.

\bibitem[Jaynes, 1985]{Jaynes:1985}
Jaynes, E.~T. (1985).
\newblock Macroscopic prediction.
\newblock In Haken, H., editor, {\em Complex Systems --- Operational Approaches
  in Neurobiology, Physics, and Computers}, pages 254--269, Berlin, Heidelberg.
  Springer Berlin Heidelberg.

\bibitem[Jeffery et~al., 2019]{JefferyEtal:2019}
Jeffery, K., Pollack, R., and Rovelli, C. (2019).
\newblock On the statistical mechanics of life: {S}chr\"odinger revisited.
\newblock {\em Entropy}, 21(12).

\bibitem[Kaiser and Schreiber, 2002]{KaiserSchreiber:2002}
Kaiser, A. and Schreiber, T. (2002).
\newblock Information transfer in continuous processes.
\newblock {\em Physica D}, 166:43--62.

\bibitem[Kim, 2006]{Kim:2006}
Kim, J. (2006).
\newblock Emergence: {C}ore ideas and issues.
\newblock {\em Synthese}, 151(3):547--559.

\bibitem[Knuth, 1985]{Knuth:1985}
Knuth, D.~E. (1985).
\newblock Semi-optimal bases for linear dependencies.
\newblock {\em Linear and Multilinear Algebra}, 17(1):1--4.

\bibitem[Kobayashi and Nomizu, 1996]{KobayashiNomizu:1996}
Kobayashi, S. and Nomizu, K. (1996).
\newblock {\em Foundations of Differential Geometry, Volume 1}.
\newblock John Wiley \& Sons, Inc.

\bibitem[Krakauer et~al., 2020]{KrakauerEtal:2020}
Krakauer, D., Bertschinger, N., Olbrich, E., Flack, J.~C., and Ay, N. (2020).
\newblock The information theory of individuality.
\newblock {\em Theory Biosci.}, 139:209--223.

\bibitem[Lancaster and Rodman, 1995]{LancasterRodman:1995}
Lancaster, P. and Rodman, L. (1995).
\newblock {\em Algebraic Riccati Equations}.
\newblock Oxford University Press, Oxford, UK.

\bibitem[L\"utkepohl, 2005]{Lutkepohl:2005}
L\"utkepohl, H. (2005).
\newblock {\em New Introduction to Multiple Time Series Analysis}.
\newblock Springer-Verlag, Berlin.

\bibitem[{Mac Lane}, 1978]{MacLane:1978}
{Mac Lane}, S. (1978).
\newblock {\em Categories for the Working Mathematician}.
\newblock Springer Science+Business Media, New York, NY, 2nd edition.

\bibitem[Mehrmann and Poloni, 2012]{MehrmannPoloni:2012}
Mehrmann, V. and Poloni, F. (2012).
\newblock Doubling algorithms with permuted {L}agrangian graph bases.
\newblock {\em SIAM J. Matrix Anal. Appl.}, 33(3):780--805.

\bibitem[Millidge, 2021]{Millidge:2021}
Millidge, B. (2021).
\newblock Towards a mathematical theory of abstraction.
\newblock {\em arXiv}, 2106.01826 [cs.AI].

\bibitem[Nelder and Mead, 1965]{NelderMead:1965}
Nelder, J.~A. and Mead, R. (1965).
\newblock A simplex method for function minimization.
\newblock {\em Comput. J.}, 7(4):308--313.

\bibitem[Neyman and Pearson, 1933]{NeymanPearson:1933}
Neyman, J. and Pearson, E.~S. (1933).
\newblock On the problem of the most efficient tests of statistical hypotheses.
\newblock {\em Phil. Trans. R. Soc. A}, 231:289--337.

\bibitem[{\O}ksendal, 2003]{Oksendal:2003}
{\O}ksendal, B. (2003).
\newblock {\em Stochastic Differential Equations: An Introduction with
  Applications}.
\newblock Springer-Verlag, Berlin.

\bibitem[Olivieri and Vares, 2005]{OlivieriVares:2005}
Olivieri, E. and Vares, M.~E. (2005).
\newblock {\em Large Deviations and Metastability}.
\newblock Cambridge University Press.

\bibitem[Palu\v{s} et~al., 2001]{Palus:2001}
Palu\v{s}, M., Kom\'arek, V., Hrn\v{c}\'i\v{r}, Z., and \v{S}t\v{e}rbov\'a, K.
  (2001).
\newblock Synchronization as adjustment of information rates: {D}etection from
  bivariate time series.
\newblock {\em Phys. Rev. E}, 63(4):046211.

\bibitem[Pearl, 1998]{Pearl:1998}
Pearl, J. (1998).
\newblock Graphical models for probabilistic and causal reasoning.
\newblock In Smets, P., editor, {\em Quantified Representation of Uncertainty
  and Imprecision}, pages 367--389. Springer Netherlands, Dordrecht.

\bibitem[Pearl, 2009]{Pearl:2009}
Pearl, J. (2009).
\newblock {\em Causality: Models, Reasoning and Inference}.
\newblock Cambridge University Press, Cambridge, UK, 2nd edition.

\bibitem[Raichle et~al., 2001]{RaichleEtal:2001}
Raichle, M.~E., MacLeod, A.~M., Snyder, A.~Z., Powers, W.~J., Gusnard, D.~A.,
  and Shulman, G.~L. (2001).
\newblock A default mode of brain function.
\newblock {\em Proc. Natl. Acad. Sci. USA}, 98(2):676--682.

\bibitem[Rechenberg, 1973]{Rechenberg:1973}
Rechenberg, I. (1973).
\newblock {\em Evolutionsstrategie -- Optimierung technischer Systeme nach
  Prinzipien der biologischen Evolution}.
\newblock Frommann-Holzboog, Stuttgart.

\bibitem[Rosas et~al., 2020a]{RosasEtal:2020}
Rosas, F.~E., Mediano, P. A.~M., Jensen, H.~J., Seth, A.~K., Barrett, A.~B.,
  Carhart-Harris, R.~L., and Bor, D. (2020a).
\newblock Reconciling emergences: {An} information-theoretic approach to
  identify causal emergence in multivariate data.
\newblock {\em PLoS Comput. Biol.}, 16(12):e1008289.

\bibitem[Rosas et~al., 2020b]{RosasEtal:2020a}
Rosas, F.~E., Mediano, P. A.~M., Rassouli, B., and Barrett, A.~B. (2020b).
\newblock An operational information decomposition via synergistic disclosure.
\newblock {\em Journal of Physics A: Mathematical and Theoretical},
  53(48):485001.

\bibitem[Rozanov, 1967]{Rozanov:1967}
Rozanov, Y.~A. (1967).
\newblock {\em Stationary Random Processes}.
\newblock Holden-Day, San Francisco.

\bibitem[Schreiber, 2000]{Schreiber:2000}
Schreiber, T. (2000).
\newblock Measuring information transfer.
\newblock {\em Phys. Rev. Lett.}, 85(2):461--4.

\bibitem[Schwefel, 1995]{Schwefel:1995}
Schwefel, H.-P. (1995).
\newblock {\em Evolution and Optimum Seeking}.
\newblock John Wiley \& Sons, Inc., New York, NY.

\bibitem[Seth, 2010]{Seth:2010}
Seth, A.~K. (2010).
\newblock Measuring autonomy and emergence via {G}ranger causality.
\newblock {\em Artificial Life}, 16(2):179--196.

\bibitem[Seth et~al., 2015]{Seth:gcneuro:2015}
Seth, A.~K., Barrett, A.~B., and Barnett, L. (2015).
\newblock Granger causality analysis in neuroscience and neuroimaging.
\newblock {\em J. Neurosci.}, 35(8):3293--3297.

\bibitem[Seth et~al., 2013]{Seth:gcfmri:2013}
Seth, A.~K., Chorley, P., and Barnett, L. (2013).
\newblock Granger causality analysis of {fMRI} {BOLD} signals is invariant to
  hemodynamic convolution but not downsampling.
\newblock {\em Neuroimage}, 65:540--555.

\bibitem[Shalizi and Moore, 2003]{ShaliziMoore:2003}
Shalizi, C.~R. and Moore, C. (2003).
\newblock What is a macrostate? {S}ubjective observations and objective
  dynamics.
\newblock {\em arXiv}, 0303625 [cond-mat.stat-mech].

\bibitem[Solo, 2016]{Solo:2016}
Solo, V. (2016).
\newblock State-space analysis of {G}ranger-{G}eweke causality measures with
  application to {fMRI}.
\newblock {\em Neural Comput.}, 28(5):914--949.

\bibitem[Spinney et~al., 2017]{SpinneyProkopenkoLizier:2017}
Spinney, R.~E., Prokopenko, M., and Lizier, J.~T. (2017).
\newblock Transfer entropy in continuous time, with applications to jump and
  neural spiking processes.
\newblock {\em Phys. Rev. E}, 95(3):032319.

\bibitem[Tononi and Sporns, 2003]{TononiSporns:2003}
Tononi, G. and Sporns, O. (2003).
\newblock Measuring information integration.
\newblock {\em BMC Neuroscience}, 4(31):1471--2202.

\bibitem[Usevich and Markovsky, 2014]{UsevichMarkovsky:2014}
Usevich, K. and Markovsky, I. (2014).
\newblock Optimization on a {G}rassmann manifold with application to system
  identification.
\newblock {\em Automatica}, 50(6):1656--1662.

\bibitem[{van Overschee} and {de Moor}, 1996]{VOandDM:1996}
{van Overschee}, P. and {de Moor}, B. L.~R. (1996).
\newblock {\em Subspace Identification for Linear Systems: Theory,
  Implementation, Applications}.
\newblock Kluwer Academic Publishers, Dordrecht, The Netherlands.

\bibitem[Wald, 1943]{Wald:1943}
Wald, A. (1943).
\newblock Tests of statistical hypotheses concerning several parameters when
  the number of observations is large.
\newblock {\em T. Am. Math. Soc.}, 54(3):426--482.

\bibitem[Weisenburger et~al., 2019]{WeisenburgerEtal:2019}
Weisenburger, S., Tejera, F., Demas, J., Chen, B., Manley, J., Sparks, F.~T.,
  {Martínez Traub}, F., Daigle, T., Zeng, H., Losonczy, A., and Vaziri, A.
  (2019).
\newblock Volumetric {Ca2+} imaging in the mouse brain using hybrid multiplexed
  sculpted light microscopy.
\newblock {\em Cell}, 177(4):1050--1066.e14.

\bibitem[Wibral et~al., 2017]{WibralEtal:2017}
Wibral, M., Priesemann, V., Kay, J.~W., Lizier, J.~T., and Phillips, W.~A.
  (2017).
\newblock Partial information decomposition as a unified approach to the
  specification of neural goal functions.
\newblock {\em Brain and Cognition}, 112:25--38.
\newblock Perspectives on Human Probabilistic Inferences and the 'Bayesian
  Brain'.

\bibitem[Wiener, 1956]{Wiener:1956}
Wiener, N. (1956).
\newblock The theory of prediction.
\newblock In Beckenbach, E.~F., editor, {\em Modern Mathematics for Engineers},
  pages 165--190. McGraw Hill, New York.

\bibitem[Wilks, 1932]{Wilks:1932}
Wilks, S.~S. (1932).
\newblock Certain generalizations in the analysis of variance.
\newblock {\em Biometrika}, 24:471--494.

\bibitem[Wilks, 1938]{Wilks:1938}
Wilks, S.~S. (1938).
\newblock The large-sample distribution of the likelihood ratio for testing
  composite hypotheses.
\newblock {\em Ann. Math. Stat.}, 6(1):60--62.

\bibitem[Williams and Beer, 2010]{WilliamsBeer:2010}
Williams, P.~L. and Beer, R.~D. (2010).
\newblock Nonnegative decomposition of multivariate information.
\newblock {\em arXiv}, 1004.2515 [cs.IT].

\bibitem[Wilson, 1972]{Wilson:1972}
Wilson, G.~T. (1972).
\newblock The factorization of matricial spectral densities.
\newblock {\em SIAM J. Appl. Math.}, 23(4):420--426.

\bibitem[Wong, 1967]{Wong:1967}
Wong, Y.-C. (1967).
\newblock Differential geometry of {G}rassmann manifolds.
\newblock {\em Proc. Natl. Acad. Sci. USA}, 57(3):589--594.

\end{thebibliography}


\end{document}